\documentclass[12pt]{article}
\usepackage{amssymb, amsmath, amsthm}
\usepackage{graphicx}
\usepackage{fullpage}
\usepackage{microtype}
\usepackage{multirow}
\usepackage{setspace}
\usepackage[super]{natbib}
\usepackage{booktabs}

\newcommand{\argmin}{\operatorname*{arg \ min}}
\newcommand{\minim}{\operatorname*{minimize}}

\usepackage{etoolbox}
\AtBeginEnvironment{algorithm}{\setstretch{1.35}}

\usepackage{fullpage}
\usepackage{hyperref}
\usepackage{amsmath}
\usepackage[table]{xcolor}
\newcolumntype{L}{>$l<$}
\theoremstyle{plain}

\newtheorem{prop}{Proposition}

\newtheorem{lemma}{Lemma}

\usepackage{algorithm,algorithmic}
\usepackage{setspace}

\begin{document}
\title{Scalable algorithms for semiparametric accelerated failure time models in high dimensions}
\author{
Piotr M. Suder and Aaron J. Molstad\footnote{Correspondence: amolstad@ufl.edu}\\
Department of Statistics and Genetics Institute\\
University of Florida, Gainesville, FL
}
\date{}

\maketitle
\begin{abstract}
Semiparametric accelerated failure time (AFT) models are a useful alternative to Cox proportional hazards models, especially when the assumption of constant hazard ratios is untenable. However, rank-based criteria for fitting AFT models are often non-differentiable, which poses a computational challenge in high-dimensional settings.
 In this article, we propose a new alternating direction method of multipliers algorithm for fitting semiparametric AFT models by minimizing a penalized rank-based loss function. Our algorithm scales well in both the number of subjects and number of predictors, and can easily accommodate a wide range of popular penalties. To improve the selection of tuning parameters, we propose a new criterion which avoids some common problems in cross-validation with censored responses. 
 Through extensive simulation studies, we show that our algorithm and software is much faster than existing methods (which can only be applied to special cases), and we show that estimators which minimize a penalized rank-based criterion often outperform alternative estimators which minimize penalized weighted least squares criteria. Application to nine cancer datasets further demonstrates that rank-based estimators of semiparametric AFT models are competitive with estimators assuming proportional hazards in high-dimensional settings, whereas weighted least squares estimators are often not.  A software package implementing the algorithm, along with a set of auxiliary functions, is available for download at \texttt{github.com/ajmolstad/penAFT}.

\end{abstract}

\onehalfspacing

\section{Introduction}
Survival analysis has applications in numerous fields of study including medicine, finance, engineering, and others.  In this article, we focus on a central task in survival analysis: modeling a time-to-event outcome as a function of a $p$-dimensional vector of predictors. Arguably, the most widely used regression model in survival analysis is the Cox proportional hazards model (henceforth, the ``Cox model''). The Cox model assumes that the ratio of hazards for any two subjects is constant across time. From a computational perspective, this assumption simplifies maximum (partial) likelihood estimation, which has led to the development of a wide range of algorithms and software packages for fitting the Cox model in both classical ($n > p$) and high-dimensional ($p \gg n$) settings.

The accelerated failure time model is an attractive alternative to the Cox model when the assumption of proportional hazards is untenable \cite{wei1992accelerated,kalbfleisch2011statistical}. The semiparametric accelerated failure time (AFT) model, which will be our focus, assumes that the failure time (e.g., survival time) for the $i$th subject, $T_i$, is the random variable 
%\vspace{-10pt}
\begin{equation}\label{eq:AFT_Model}
 \log T_i = \beta_*^\top x_i + \epsilon_i, \quad i\in \{1, \dots, n\}
 %\vspace{-10pt}
 \end{equation}
where $x_i \in \mathbb{R}^{p}$ is the vector of predictors, $\beta_* \in \mathbb{R}^p$ is a vector of unknown regression coefficients, and $\epsilon_1, \dots, \epsilon_n$ are independent and identically distributed errors with an unspecified distribution. In practice we may not observe realizations of all $T_i$. Instead, we observe realizations of $Y_i = \min(T_i, C_i)$ where $C_i$ is a random censoring variable for the $i$th subject which is independent of $T_i$ for $i \in \{1, \dots, n\}$. Thus, the data we use to fit the model in \eqref{eq:AFT_Model} are $\left\{(y_1, x_1, \delta_1), \dots, (y_n, x_n, \delta_n)\right\}$ where $y_i$ is a realization of $Y_i$ and $\delta_i = \mathbf{1}(y_i = t_i)$ is the censoring indicator where $t_i$ is the (possibly unobserved) realization of $T_i$ for $i\in \{1, \dots, n\}$. 

There are numerous approaches to fit the model in \eqref{eq:AFT_Model}. To simplify computation, AFT models are sometimes fit under a parametric assumption on the distribution of the $\epsilon_i$'s. However, parametric restrictions reduce the flexibility of AFT models and thus make them a less attractive alternative to the Cox model. To avoid parametric assumptions, it is common to estimate $\beta_*$ using weighted least squares \cite{zhou1992m,stute1993consistent,stute1996distributional,huang2006regularized} or rank-based criteria \cite{prentice1978linear,tsiatis1990estimating}. To use weighted least squares, weights for censored failure times are reassigned to the observed failure times. If no censoring has occurred, this approach is equivalent to using the unweighted least squares estimator for $\beta_*$. It it well understood that if the distribution of the $\epsilon_i$'s is asymmetric or  heavy-tailed, the least squares estimator may perform poorly. Thus, rank-based estimators are often preferable. However, because rank-based estimators are often difficult to compute, weighted least squares estimators are frequently used in practice despite their potential deficiencies. Later, we will show that estimators which minimize rank-based criteria outperform weighted least squares estimators under a variety of data generating models. 

One of the more widely used rank-based estimation criteria is the so-called Gehan loss function
%\vspace{-10pt}
\begin{equation} \label{eq:gehan_loss}
\frac{1}{n^2} \sum_{i=1}^n \sum_{j=1}^n \delta_i \{ e_i(\beta) - e_j(\beta) \}^{-}, \quad e_i(\beta) = \log y_i - \beta^\top x_i, \quad i \in \{1, \dots, n\},
%\vspace{-10pt}
\end{equation}
where we define $a^{-} = \max(-a, 0).$
This loss was originally inspired by Tsiatis \cite{tsiatis1990estimating}, who proposed to estimate $\beta_*$ using a weighted log-rank estimating equation. With weights from Gehan,\cite{gehan1965generalized} Tsiatis's weighted log-rank estimating function is monotone \citep{fygenson1994monotone} and is a selection of the subdifferential of \eqref{eq:gehan_loss}.  Thus \eqref{eq:gehan_loss}, which is convex, is a well-motivated choice of loss function for estimating $\beta_*.$

In modern survival analyses -- especially those involving genetic or genomic data -- it is often the case that $p \gg n$. In such settings, it is common to use a regularized estimator of $\beta_*.$ The estimator we focus on in this work is the regularized Gehan estimator
%\vspace{-10pt}
\begin{equation} \label{eq:penalizedGehan}
 \argmin_{{\beta} \in \mathbb{R}^p}  \left\{ \frac{1}{n^2} \sum_{i=1}^n \sum_{j=1}^n \delta_i \{ e_i(\beta) - e_j(\beta)\}^{-} + \lambda g({\beta})\right\}
 %\vspace{-10pt}
\end{equation}
where $g: \mathbb{R}^p \to \mathbb{R}_+$ is assumed to be a convex penalty function and $\lambda > 0$ is a user specified tuning parameter. When $g$ is convex, the objective function in \eqref{eq:penalizedGehan} is convex. The function $g$ could be, for example, the $L_1$-norm (i.e., the lasso penalty). With this choice of $g$, for sufficiently large values of the tuning parameter $\lambda$, many entries of \eqref{eq:penalizedGehan} will be equal to zero. In high-dimensional settings, this can lead to improved estimation accuracy and interpretability.

While the $L_1$-penalized version of \eqref{eq:penalizedGehan} has appeared in the literature \citep{johnson2009rank,cai2009regularized}, the matter of computing \eqref{eq:penalizedGehan} is largely unresolved, even in this special case \cite{chung2013tutorial}. Although \eqref{eq:penalizedGehan} is the solution to a convex optimization problem, the objective function is (depending on $g$) often the sum of two non-differentiable functions. Thus, standard first and second order methods cannot be applied, so many ``off-the-shelf'' solvers are not able to compute \eqref{eq:penalizedGehan} efficiently. Approximations to \eqref{eq:penalizedGehan}, which we will discuss in a later section, lead to optimization problems which are arguably no easier to solve. Needless to say, there exist no publicly available software packages for solving \eqref{eq:penalizedGehan} beyond the $L_1$-penalized case. Cox model analogs of \eqref{eq:penalizedGehan}, on the other hand, have numerous fast and easy-to-use software packages which can handle a wide variety of penalties $g$, e.g., \texttt{grpreg} \citep{grpreg} accommodates the group-lasso penalty and \texttt{glmnet} \citep{simon2011regularization} the elastic net penalty.
% In the era of high-dimensional multi-omic data, which are often used in survival analyses, it is essential that a wide range of candidate models, like the semiparametric accelerated failure time model, can be implemented by practicioners. 

In this article, we propose a unified algorithm for fitting the semiparametric accelerated failure time model using \eqref{eq:penalizedGehan} that can be applied to a broad class of penalty functions $g$ and scales efficiently in $n$ and $p$.  Our algorithm performs favorably compared to existing approaches for computing the $L_1$-penalized version of \eqref{eq:penalizedGehan}. Moreover, we perform a comprehensive comparison of penalized weighted least squares estimators to \eqref{eq:penalizedGehan} in high-dimensional settings and show that the rank-based estimators perform better under various data generating models. Later, we also show that penalized rank-based estimators are competitive with the estimators assuming proportional hazards in nine cancer datasets, whereas the weighted least squares estimators are not. An R package implementing our method, along with a set of auxiliary functions for prediction, cross-validation, and visualization, is available for download at \url{https://github.com/ajmolstad/penAFT}. 

Before describing our algorithm, we first discuss the data analysis which motivated our work and then describe existing methods for solving special cases of \eqref{eq:penalizedGehan}.

\subsection{Motivating pathway-based analysis of KIRC dataset}\label{sec:pathway_survival}
The work in this article was motivated in part by a pathway-based survival analysis of a kidney renal clear cell carcinoma (KIRC) survival dataset collected by The Cancer Genome Atlas project \cite{weinstein2013cancer} (TCGA, https://portal.gdc.cancer.gov/). The goal was to analyze the effect of gene expression on survival while treating genes belonging to a set of biologically relevant pathways as groups \cite{10.1093/bioinformatics/btz446,Molstad2019Gaussian}. In this case, the estimator we would like to use employs a variation of the sparse group lasso penalty \cite{simon2013sparse} 
%\vspace{-10pt}
\begin{equation} \label{eq:sparseGroupLasso}
\argmin_{\beta \in \mathbb{R}^p}  \left\{ \frac{1}{n^2} \sum_{i=1}^n \sum_{j=1}^n \delta_i \{ e_i(\beta) - e_j(\beta) \}^{-} + \lambda \alpha \|w \circ \beta\|_1 + \lambda (1- \alpha) \sum_{g=1}^G v_g \|\beta_{\mathcal{G}_g}\|_2 \right\}
%\vspace{-10pt}
\end{equation}
where $\lambda > 0$ and $\alpha \in [0,1]$ are tuning parameters; $\left\{\mathcal{G}_1, \dots, \mathcal{G}_G\right\}$ is a $G$ element partition of $\left\{1, \dots, p\right\}$; ${\beta}_{\mathcal{G}_g}$ is the subvector of ${\beta}$ whose components are indexed by $\mathcal{G}_g$; $w \in \mathbb{R}^p$ and the $v_g \in \mathbb{R}$ $(g \in \{1, \dots, G\})$ 
are non-negative weights; $\|\cdot\|_1$ and $\|\cdot\|_2$ denote the $L_1$ and $L_2$ (Euclidean) norms, respectively; and $\circ$ denotes the elementwise product. In this context, $\mathcal{G}_g$ denotes the set of genes belonging to the $g$th pathway and $v_g$ 
is a user-specified weight corresponding to all coefficients from the $g$th pathway. When estimating $\beta_*$ with \eqref{eq:sparseGroupLasso}, as $\lambda (1-\alpha)$ is increased, some pathways will have all their estimated coefficients equal to zero. As $\lambda \alpha$ is increased, pathways with some nonzero coefficient estimates will have a subset of coefficients equal to zero. Thus, we can think of the estimator in \eqref{eq:sparseGroupLasso} as performing ``bi-level'' variable selection \citep{breheny2009penalized} in the sense that it can select both pathways and specific genes within pathways. With fitted models that can be interpreted in this way, the molecular mechanisms underlying survival can be more precisely characterized in terms of the the known biological functions of gene pathways. 

While the estimator in \eqref{eq:sparseGroupLasso} is well-motivated, to the best of our knowledge, it has not been used in the literature. We suspect this is due to the fact that existing computational approaches for computing \eqref{eq:penalizedGehan}, which we describe in the next section, cannot be easily modified to solve \eqref{eq:sparseGroupLasso}. Our algorithm and software, in contrast, can easily handle problems like \eqref{eq:sparseGroupLasso}, which makes a much wider range of estimators accessible to practitioners.

\section{Existing approaches}\label{eq:ExistingMethods}
There exist numerous approaches for solving special cases of \eqref{eq:penalizedGehan}. We discuss two in depth here and we compare these to our algorithm in a later section.  For a more thorough review of existing computational methods in the unpenalized setting, we refer readers to the tutorial of Chung et al \cite{chung2013tutorial}. 

The main approach for solving the $L_1$-penalized version of \eqref{eq:penalizedGehan} formulates the optimization problem as a linear program\cite{chung2013tutorial}. The formulation of the linear program described in Cai et al \cite{cai2009regularized} is
%%\vspace{-10pt}
\begin{equation}\label{eq:linProg}
\minim_{\beta \in \mathbb{R}^p}  \left\{ \frac{1}{n^2} \sum_{i=1}^n \sum_{j=1}^n \delta_i \ddot{e}_{i,j}(\beta)\right\} ~~~~\text{ subject to }~~e_{i,j}(\beta) = \log y_i - \log y_j - \beta^\top (x_i - x_j),
%%\vspace{-10pt}
\end{equation}
$$ e_{i,j}(\beta) =\dot{e}_{i,j}(\beta) - \ddot{e}_{i,j}(\beta),~~~ \dot{e}_{i,j}(\beta) \geq 0,~~ \ddot{e}_{i,j}(\beta) \geq 0, ~~~ (i,j) \in[n] \times [n]$$
$$ \beta_k = \dot{\beta}_k - \ddot{\beta}_k, ~~~\sum_{k=1}^p (\dot{\beta}_k + \ddot{\beta}_k) \leq \tilde{\lambda}, ~~~\dot{\beta}_k \geq 0,~~ \ddot{\beta}_k \geq 0, ~~~k \in [p]$$
where $\tilde{\lambda} > 0$ is a tuning parameter (analogous to  $\lambda$ in \eqref{eq:penalizedGehan}) and by definition, $[n]= \{1, 2, \dots, n\}$ for any $n \in \mathbb{N}$. Here, the notation $\dot{b}$ and $\ddot{b}$ is used to represent the positive and negative parts of $b$, respectively, so that we can write $b = \dot{b} - \ddot{b}$ for any $b \in \mathbb{R}$ where $\dot{b} \geq 0$ and $\ddot{b} \geq 0$. The linear program in \eqref{eq:linProg} can be solved using simplex or interior point methods, both of which can require prohibitively long computing times when $n$ or $p$ is large since there are $O(n^2 + p)$ constraints. This becomes especially problematic since in practice, one often needs to solve \eqref{eq:penalizedGehan} over a grid of candidate tuning parameters multiple times to perform cross-validation.

To partially alleviate this issue, Cai et al \cite{cai2009regularized} derived an approach which computes the solution to the linear program \eqref{eq:linProg} along a path of increasing values for $\tilde{\lambda}$ (i.e., computes the ``solution path''). Their approach relies on the fact that the solution path is piecewise linear in $\tilde{\lambda}$. While their algorithm can be faster than naively employing ``off-the-shelf'' linear programming methods to solve \eqref{eq:linProg} and yields extremely accurate solutions, there are often many kinks in the path wherein no new coefficients become non-zero. Thus, one must compute $\beta$ at many candidate tuning parameter values $\tilde{\lambda}$ to reach even a moderately non-sparse model. Moreover, computing each new point along the solution path is itself computationally burdensome and thus, this approach does not scale to truly high-dimensional settings. For example, in their simulation studies, Cai et al \cite{cai2009regularized} considered dimensions $(n,p) = (100, 9)$ and $(n,p) = (50,50).$

Taking a different approach than Cai et al \cite{cai2009regularized}, Johnson \cite{johnson2008estimation,johnson2009rank} relied on a reformulation to \eqref{eq:penalizedGehan}, which was suggested in Jin et al \cite{jin2003rank}. They define the function 
$$ h_M(\beta) = \frac{1}{n^2} \sum_{i=1}^n \sum_{j=1}^n \delta_i |e_i(\beta) - e_j(\beta)| + \frac{1}{n^2} \left| M - \beta^\top \sum_{i=1}^n \sum_{j=1}^n \delta_i (x_j - x_i) \right|,$$
and use that the argument minimizing $h_M(\beta) + \lambda g(\beta)$ is equivalent to \eqref{eq:penalizedGehan} when $M$ is taken to be a sufficiently large constant (e.g., $M = n^2 10^4 $). This is especially convenient when $g$ is the $L_1$-norm because the resulting optimization problem can be expressed as a least absolute deviations optimization problem
$\argmin_{\beta \in \mathbb{R}^{p}} \{ \|w - \tilde{X}\beta\|_1\}$
for a $w \in \mathbb{R}^{n \sum_{i=1}^n \delta_i + p + 1}$ and $\tilde{X} \in \mathbb{R}^{(n \sum_{i=1}^n \delta_i + p + 1) \times p}$ constructed from the $y_i$, $x_i$, $\delta_i$, $M$, and $\lambda$.  Johnson \cite{johnson2009rank} solved this problem using the package \texttt{quantreg} in R, which uses an interior point method for solving the corresponding linear program. This formulation is very convenient, but as discussed in Chung et al \cite{chung2013tutorial}, can require long computing times when $n$ or $p$ are large. In the time since Johnson \cite{johnson2009rank} was published, new R packages have been developed for solving regularized least absolute deviations problems like $\argmin_{\beta \in \mathbb{R}^p} h_M(\beta) + \lambda g(\beta)$, e.g., \texttt{hqreg}\cite{yi2017semismooth}. However, we found this software could be both slow and inaccurate in certain settings: see Section \ref{subsec:compare_to_existing} for further details.
 
The path-based approach of Cai et al \cite{cai2009regularized} and interior point approach of Johnson \cite{johnson2009rank} are two specialized methods for solving \eqref{eq:penalizedGehan} with $g$ being the $L_1$-norm.  To solve \eqref{eq:penalizedGehan} with more general penalty functions, Chung et al \cite{chung2013tutorial} suggested replacing both terms in \eqref{eq:penalizedGehan} with smooth approximations. This raises new issues: first, smooth approximations to sparsity inducing penalties often lead to non-sparse solutions. Second, this again requires the development of a new specialized algorithmic approach for any choice of $g.$ The ideal resolution would be an algorithm which solves \eqref{eq:penalizedGehan} directly and can be easily modified to handle a large class of penalty functions $g$. The objective of this work is to derive such an algorithm, and to provide simple, modular software implementing the algorithm. 

\section{Prox-linear ADMM algorithm}\label{sec:ADMM}
%\vspace{-10pt}
\subsection{Overview of ADMM}\label{sec:ADMM_overview}
In this section, we will propose a variation of the alternating direction method of multipliers (ADMM) algorithm \citep{boyd2011distributed, deng2016global} for solving \eqref{eq:penalizedGehan} under a variety of penalties $g$. Loosely speaking, the ADMM algorithm is an efficient algorithm for solving convex constrained optimization problems of the form 
% $$\minim_\beta f(A\beta - B) + g(\beta)$$
% for some matrices or vectors $A$ and $B$. 
% It does so by first converting the unconstrained problem to a constrained problem, e.g., 
\begin{equation}\label{eq:constrained_ADMM}
\minim_{\theta \in \mathbb{R}^s, \beta \in \mathbb{R}^p}~ \{ f(\theta) + \lambda g(\beta) \}~~~\text{subject to} ~~~A\theta + B\beta = C
\end{equation}
for convex functions $f$ and $g$; and some fixed $A \in \mathbb{R}^{a \times s}, B \in \mathbb{R}^{a \times p}$ and $C \in \mathbb{R}^a.$ 
By exploiting that for $\rho > 0$, 
 \begin{equation}\label{eq:augmented_constrained}
 \minim_{\theta \in \mathbb{R}^s, \beta \in \mathbb{R}^p}  ~\left\{ f(\theta) + \lambda g(\beta) + \frac{\rho}{2} \|A\theta + B\beta - C\|_2^2 \right\}~~~\text{subject to} ~~~ A\theta + B\beta = C
 \end{equation}
is an equivalent problem (since the quadratic term is zero on the set of $(\theta, \beta)$ such that $A\theta + B\beta = C$), the ADMM algorithm solves \eqref{eq:augmented_constrained} using a variation of the augmented Lagrangian method (also known as the method of multipliers). 
In brief, the augmented Lagrangian method introduces Lagrangian dual variable $\Gamma \in \mathbb{R}^a$ and updates $(\theta, \beta)$ and $\Gamma$ from $(t-1)$th to $(t)$th iterates using 
\begin{align}  
(\theta^{(t)}, \beta^{(t)}) & = \argmin_{\theta \in \mathbb{R}^s, \beta \in \mathbb{R}^p} \left\{f(\theta) + \lambda g(\beta) + \frac{\rho}{2} \|A\theta + B\beta - C\|_2^2 + (A \theta + B \beta - C)^\top {\Gamma^{(t-1)}}\right\},\label{eq:joint_update}\\
\Gamma^{(t)} & =  \Gamma^{(t-1)} + \rho (A\theta^{(t)} + B\beta^{(t)} - C). \notag
\end{align}
The ADMM algorithm modifies \eqref{eq:joint_update} by updating $\theta$ and $\beta$ separately so that after some algebra, 
\begin{align}  
\beta^{(t)} &=  \argmin_{\beta \in \mathbb{R}^p} \left\{\lambda g(\beta) + \frac{\rho}{2} \|A\theta^{(t-1)} + \rho^{-1}\Gamma^{(t-1)} - C + B\beta \|_2^2\right\}, \label{eq:b1} \\
\theta^{(t)} &=  \argmin_{\theta \in \mathbb{R}^s} \left\{f(\theta) + \frac{\rho}{2} \|A\theta + \rho^{-1}\Gamma^{(t-1)} - C + B\beta^{(t)} \|_2^2\right\},\label{eq:t1}
\end{align}
which serves to decouple the functions $f$ and $g$. This decoupling can greatly simplify the updates of $\theta$ and $\beta$ relative to the joint update of $(\theta, \beta)$ in the augmented Lagrangian method. Because of the quadratic (augmentation) term introduced in \eqref{eq:augmented_constrained}, both $\beta^{(t)}$ and $\theta^{(t)}$ updates in \eqref{eq:b1} and \eqref{eq:t1} can be recognized as penalized least squares problems. When either $A$ or $B$ (or both) are identity matrices, as is common in many applications, these updates simplify to the so-called proximal operators of the functions $\lambda g$ and $f$. The proximal operator of a function $h:\mathcal{X} \to \mathbb{R}$ is defined as 
%\vspace{-15pt}
$$ {\rm Prox}_h(x) = \argmin_{y \in \mathcal{X}} \left\{\frac{1}{2}\|x - y\|_2^2 + h(y)\right\}%\vspace{-10pt}
.$$ When $h$ is a proper and lower semi-continuous convex function, its proximal operator is unique. For many popular convex penalties $ \lambda g$, the proximal operator can be solved in closed form. 

To make matters concrete, in later sections we will focus on two penalties: the weighted elastic net \citep{zou2009adaptive} and the weighted sparse group lasso \citep{simon2013sparse}. We define these penalties and give the closed form of their proximal operators in Table \ref{tab:Penalties}. In the derivation of our algorithm, we leave $g$ arbitrary to demonstrate how this algorithm could be applied in other settings. 

\begin{table}
\centering
\fbox{
\begin{tabular}{cll}
$g(\beta)$ ~~$(\alpha \in [0,1])$ & $\bar{\beta} = {\rm Prox}_{\lambda g}(\beta)$ & \\
\hline
Sparse group lasso&  $(1)~~ \dot{\beta}_{\mathcal{G}_l} = \max(|\beta_{\mathcal{G}_l}| - w_{\mathcal{G}_l} \alpha\lambda, 0){\rm sign}(\beta_{\mathcal{G}_l})$  & \multirow{2}{*}{$l \in [G]$}\\
$\alpha \|w \circ \beta\|_1 + (1-\alpha)\sum_{l=1}^G v_l \|\beta_{\mathcal{G}_l}\|_2$ &   $(2)~ \bar{\beta}_{\mathcal{G}_l} = \max(\|\dot{\beta}_{\mathcal{G}_l}\|_2 - v_l (1-\alpha)\lambda, 0) \dot{\beta}_{\mathcal{G}_l}/\|\dot{\beta}_{\mathcal{G}_l}\|_2$  &\\
\\
Elastic net &  $(1) ~~ \dot{\beta} = \max(|\beta| - w \alpha \lambda, 0){\rm sign}(\beta)$ & \\
$\alpha \|w \circ \beta\|_1 + \frac{(1-\alpha)}{2} \|\beta\|_2^2$ & $(2) ~ \bar{\beta} = \dot{\beta}/(1 + (1 - \alpha)\lambda)$& 
\end{tabular}
}
\caption{The two penalty functions implemented in our software \texttt{penAFT} and their corresponding proximal operators, which are computed in two closed-form steps. }\label{tab:Penalties}
\end{table}

In the next subsection, we show that \eqref{eq:penalizedGehan} can be expressed as \eqref{eq:constrained_ADMM} for a particular function $f$ with $A$ being an identity matrix. Then, we derive a closed form expression for the corresponding $\theta$ update; and devise an approximation to the $\beta$ update which involves only the proximal operator of $\lambda g.$

% Thus, the efficiency of ADMM often hinges on the ability to compute the proximal operator of the functions $f$ and $g$ efficiently \citep{parikh2014proximal,polson2015proximal}.

% \begin{table}
% \centering
% \fbox{
% \begin{tabular}{cll}
% $g(\beta)$ ~~$(\alpha \in [0,1])$ & $\bar{\beta} = {\rm Prox}_{\lambda g}(\beta)$ & \\
% \hline
% Sparse group lasso&  $(1)~~ \dot{\beta}_{\mathcal{G}_l} = \max(|\beta_{\mathcal{G}_l}| - w_{\mathcal{G}_l} \alpha\lambda, 0){\rm sign}(\beta_{\mathcal{G}_l})$  & \multirow{2}{*}{$l \in [G]$}\\
% $\alpha \|w \circ \beta\|_1 + (1-\alpha)\sum_{l=1}^G v_l \|\beta_{\mathcal{G}_l}\|_2$ &   $(2)~ \bar{\beta}_{\mathcal{G}_l} = \max(\|\dot{\beta}_{\mathcal{G}_l}\|_2 - v_l (1-\alpha)\lambda, 0) \dot{\beta}_{\mathcal{G}_l}$  &\\
% \\
% Elastic net &  $(1) ~~ \dot{\beta} = \max(|\beta| - w \alpha \lambda, 0){\rm sign}(\beta)$ & \\
% $\alpha \|w \circ \beta\|_1 + \frac{(1-\alpha)}{2} \|\beta\|_2^2$ & $(2) ~ \bar{\beta} = \dot{\beta}/(1 + (1 - \alpha)\lambda)$& 
% \end{tabular}
% }
% \caption{The two penalty functions implemented in our software \texttt{penAFT} and their corresponding proximal operators, which are computed in two closed-form steps. }\label{tab:Penalties}
% \end{table}

\subsection{Formulation and updating equations}
As mentioned, we use a variation of the ADMM algorithm to solve \eqref{eq:penalizedGehan}. 
% Briefly, this approach  (i) introduces a new set of optimization variables corresponding to all (necessary) pairwise differences of the $y_i - \beta'x_i$, (ii) uses these to formulate a constrained version of \eqref{eq:penalizedGehan}, and (iii) solves this constrained problem using an approach combining dual ascent and the method of multipliers. 
To begin, we rewrite the optimization problem from \eqref{eq:penalizedGehan} as a constrained problem. Naively, we may write the optimization for \eqref{eq:penalizedGehan} as 
%\vspace{-15pt}
\begin{equation}\label{eq:constrainedProb}
\minim_{\{\theta_{i,j}\}_{(i,j) \in [n]\times[n]}, {\beta} \in \mathbb{R}^p} \left\{ \frac{1}{n^2}\sum_{i=1}^n \sum_{j=1}^n\delta_i (\theta_{i,j})^{-} + \lambda g(\beta) \right\}\end{equation} 
$$\text{subject to}~~~\theta_{i,j} = \log y_i  - \log y_j - \beta^\top (x_i - x_j), ~~ (i,j) \in [n] \times [n].$$
While it is clear that the solution to \eqref{eq:constrainedProb} is the solution to \eqref{eq:penalizedGehan}, there are many redundancies in the $n^2$ variables $\theta_{i,j}$, which would impose a substantial burden on memory and storage. Instead, we use that $\theta_{i,j} = -\theta_{j,i}$ and the fact that if $\delta_i = 0$ and $\delta_j = 0$, the value of $\theta_{i,j}$ does not affect \eqref{eq:constrainedProb} to reduce the number of constraints. Thus, letting 
$\mathcal{D} = \left\{(i,j): 2 \leq j \leq n,  1 \leq i < j, \delta_i + \delta_j \geq 1 \right\},$ we can rewrite \eqref{eq:constrainedProb} with fewer constraints as
%\vspace{-10pt}
\begin{equation}\label{eq:Efficient_constrainedProb}
\minim_{\beta \in \mathbb{R}^p, \{ \theta_{i,j}\}_{(i,j) \in \mathcal{D}}} \left[ \frac{1}{n^2}   \sum_{(i,j) \in \mathcal{D}} \left\{ \delta_i ( \theta_{i,j})^{-} + \delta_j (-\theta_{i,j})^{-} \right\} + \lambda g(\beta) \right]
\end{equation}
$$~~~\text{subject to}~~~  \theta_{i,j} = \log y_i  - \log y_j - \beta^\top (x_i - x_j), ~~~ (i,j) \in \mathcal{D}.$$
%\vspace{-10pt}
To simplify notation, let $\theta \in \mathbb{R}^{|\mathcal{D}|}$ denote the collection of all $\theta_{i,j}$ for $(i,j) \in \mathcal{D}$ where $|\mathcal{D}|$ denotes the cardinality of $\mathcal{D}$, let $y = (y_1, y_2, \dots, y_n)^\top \in \mathbb{R}^n$, and let $X = (x_1, \dots, x_n)^\top \in \mathbb{R}^{n \times p}$. Then, we can  define
$ f_{\mathcal{D}}(\theta) = n^{-2} \sum_{(i,j) \in \mathcal{D}} \left\{ \delta_i ( \theta_{i,j})^{-} + \delta_j (-\theta_{i,j})^{-} \right\}$
so that we can write the constrained optimization problem from \eqref{eq:Efficient_constrainedProb} as
%\vspace{-10pt}
\begin{equation}\label{eq:Eff_ConstrainedProb}
 \minim_{\beta \in \mathbb{R}^p, \theta \in \mathbb{R}^{|\mathcal{D}|}} \left\{ f_{\mathcal{D}}(\theta) + \lambda g (\beta) \right\}~~~\text{subject to} ~~  \theta = {\rm P}_\mathcal{D} (\log y - X \beta),
 %\vspace{-10pt}
 \end{equation}
where ${\rm P}_\mathcal{D} \in \mathbb{R}^{|\mathcal{D}| \times n}$ is a matrix whose rows have $i$th element equal to one, $j$th element equal to negative one, and zeros in all other elements for each pair $(i,j) \in \mathcal{D}$. The ADMM algorithm can be then used to solve \eqref{eq:Eff_ConstrainedProb}. The updating equations for ADMM can be written in terms of the augmented Lagrangian for the constrained problem in \eqref{eq:Eff_ConstrainedProb}, which is 
%\vspace{-10pt}
$$
\mathcal{F}_\rho(\theta, \beta, \Gamma)  = f_{\mathcal{D}}(\theta) + \lambda g(\beta) + \Gamma^\top\{ \theta - {\rm P}_\mathcal{D} (\log y - X\beta)\} + \frac{\rho}{2}\|\theta - {\rm P}_\mathcal{D} (\log y - X\beta)\|_2^2,
%\vspace{-10pt}
$$
where the $\Gamma \in \mathbb{R}^{|\mathcal{D}|}$ is a Lagrangian dual variable and $\rho>0$ is a step size. A variation of the ADMM algorithm as discussed in the previous section (e.g., see Algorithm 2 of Deng and Yin\cite{deng2016global}) has $(t)$th iterates defined as
%\vspace{-15pt}
\begin{align}
    \beta^{(t)} &= \argmin_{\beta \in \mathbb{R}^{p}} \mathcal{F}_\rho(\theta^{(t-1)}, \beta, \Gamma^{(t-1)})\label{eq:beta_update}\\
        \theta^{(t)} & = \argmin_{\theta \in \mathbb{R}^{|\mathcal{D}|}} \mathcal{F}_\rho(\theta, \beta^{(t)}, \Gamma^{(t-1)}) \label{eq:theta_update}\\
    \Gamma^{(t)} &= \Gamma^{(t-1)} +  \tau\rho \{ \theta^{(t)} - {\rm P}_\mathcal{D} (\log y - X \beta^{(t)}) \} \notag
\end{align}
    %\vspace{-40pt}

\noindent 
where $\tau \in (0, (1 + \sqrt{5})/2)$ is a relaxation factor.  See, for example, Theorem 2.2 of Deng and Yin \cite{deng2016global} for more on $\tau.$
Obtaining the $(t)$th iterate of the ADMM algorithm requires solving the optimization problems in \eqref{eq:beta_update} and \eqref{eq:theta_update}. %We will first show that \eqref{eq:theta_update} can be computed in closed form. Then, we will propose a closed-form approximation to $\s^{(t)}$. 

First, we focus on \eqref{eq:theta_update}. Let $\tilde{\delta} \in \mathbb{R}^{|\mathcal{D}| \times 2}$ be a matrix with rows $(\delta_i, \delta_j)$ for each $(i,j) \in \mathcal{D}$, and let $\tilde{\delta}_{k,l} \in \mathbb{R}$ denote entry in the $k$th row and $l$th column of $\tilde{\delta}$. Note that the rows of $\tilde{\delta}$, ${\rm P}_\mathcal{D}$, and $\theta$ all correspond to the same pairs $(i,j) \in \mathcal{D}$.  With $\tilde{\delta}$ defined, we can solve \eqref{eq:theta_update} using the following lemma. 
\begin{lemma}\label{prop_theta}
Let $\phi^{(t)} = {\rm P}_\mathcal{D} (\log y - X \beta^{(t)}) - \rho^{-1}\Gamma^{(t-1)}$. For $k \in [|\mathcal{D}|]$, the $k$th element of $\theta^{(t)}$, $\theta^{(t)}_k$, is given by
$$\theta^{(t)}_k = \left\{ \begin{array}{ll}
\phi^{(t)}_k - \tilde{\delta}_{k,2}/\rho n^2 &:  \phi^{(t)}_k  > \tilde{\delta}_{k,2}/\rho n^2\\
\phi^{(t)}_k + \tilde{\delta}_{k,1}/\rho n^2 &:  \phi^{(t)}_k  < - \tilde{\delta}_{k,1}/\rho n^2\\
0 & :\text{otherwise}
\end{array}\right.$$
\end{lemma}

The result of Lemma \ref{prop_theta} reveals that we can efficiently update $\theta$ in closed-form and in parallel.  A proof of Lemma \ref{prop_theta} can be found in the Supplementary Materials.

Next, we focus on the update for $\beta$ in \eqref{eq:beta_update}.
Notice that computing $\argmin_{\beta \in \mathbb{R}^{p}} \mathcal{F}_\rho(\theta^{(t-1)}, {\beta}, {\Gamma}^{(t-1)})$ may be prohibitively expensive as this requires solving a penalized least squares problem
%\vspace{-10pt}
$$\argmin_{\beta \in \mathbb{R}^{p}}  \left\{ \lambda g(\beta) + \frac{\rho}{2}\|\theta^{(t-1)} + \rho^{-1} \Gamma^{(t-1)} - {\rm P}_\mathcal{D} (\log y - X\beta)\|_2^2 \right\}.
$$
Repeating this at each iteration may be too costly to be practical when $p$ is large.
Instead, we approximate \eqref{eq:beta_update} by minimizing a quadratic approximation to $\mathcal{F}_\rho(\theta^{(t-1)}, \beta, \Gamma^{(t-1)})$ constructed at the previous iterate $\beta^{(t-1)}$. Specifically, we add a quadratic expression to the objective function to define $h_{\eta, \rho}(\cdot \mid \beta^{(t-1)},\Gamma^{(t-1)}, \theta^{(t-1)})$ as
%\vspace{-15pt}
\begin{align*}
h_{\eta, \rho}(\beta \mid \beta^{(t-1)},\Gamma^{(t-1)}, \theta^{(t-1)}) = & \left\{ \lambda g({\beta}) + \frac{\rho}{2}\|{\theta}^{(t-1)} + \rho^{-1} {\Gamma}^{(t-1)} - {\rm P}_\mathcal{D}(\log y -  {X}{\beta})\|_2^2 \right.\\
&     ~~~~~~~~~~~~~~~~~~~~~~~~~~~~~~~~~ \left.  + \frac{\rho}{2} ({\beta} - {\beta}^{(t-1)})^\top{Q}_\eta({\beta} - {\beta}^{(t-1)}) \right\},
\end{align*}
%\vspace{-40pt}

\noindent where ${Q}_\eta = \eta I_p - X^\top{\rm P}_\mathcal{D}^\top{\rm P}_\mathcal{D}X$ with $\eta \in \mathbb{R}$ chosen so that ${Q}_\eta$ is non-negative definite. To simplify matters, define $\eta$ to be the largest eigenvalue of $X^\top{\rm P}_\mathcal{D}^\top{\rm P}_\mathcal{D}X$. Then, we replace \eqref{eq:beta_update} with 
$ \argmin_{{\beta} \in \mathbb{R}^p}\{ h_{\eta, \rho}({\beta} \mid {\beta}^{(t-1)},{\Gamma}^{(t-1)}, {\theta}^{(t-1)}) \}.$
After some algebra (see Section 1.2 of the Supplementary Material), this simplifies to
% $$  \argmin_{{\beta} \in \mathbb{R}^p} \left\{ \frac{1}{2}\left\Vert {\beta} - \frac{1}{\eta}{X}'{\rm P}_\mathcal{D}'\{{\rm P}_\mathcal{D} ({y} - {\rm P}_\mathcal{D}{X}{\beta}^{(t-1)})  - {\theta}^{(t-1)} - \rho^{-1}{\Gamma}^{(t-1)} \} - {\beta}^{(t-1)}\right\Vert_2^2 + \frac{\lambda}{\rho\eta}g({\beta}) \right\},$$
% which is the proximal operator of $(\rho\eta)^{-1}\lambda g$ evaluated at $\eta^{-1}{X}'{\rm P}_\mathcal{D}'\{{\rm P}_\mathcal{D} ({y} - {\rm P}_\mathcal{D}{X}{\beta}^{(t-1)})  - {\theta}^{(t-1)} - \rho^{-1}{\Gamma}^{(t-1)} \} + {\beta}^{(t-1)}$, i.e., we can write
% \begin{equation} \label{eq:beta_update_approx}
% {\beta}^{(t)} = {\rm Prox}_{\frac{\lambda}{\rho\eta} g}\left(\frac{1}{\eta}{X}'{\rm P}_\mathcal{D}'\{{\rm P}_\mathcal{D} ({y} - {\rm P}_\mathcal{D}{X}{\beta}^{(t-1)})  - {\theta}^{(t-1)} - \rho^{-1}{\Gamma}^{(t-1)} \} + {\beta}^{(t-1)}\right).
% \end{equation}
%\vspace{-10pt}
\begin{equation} \label{eq:beta_update_approx}
{\beta}^{(t)} = {\rm Prox}_{(\lambda/\rho\eta) g}\left[\frac{1}{\eta}{X}^\top{\rm P}_\mathcal{D}^\top\{{\rm P}_\mathcal{D} (\log y - {X}{\beta}^{(t-1)})  - {\theta}^{(t-1)} - \rho^{-1}{\Gamma}^{(t-1)} \} + {\beta}^{(t-1)}\right].
%\vspace{-10pt}
\end{equation}
which is the proximal operator of $(\lambda/\rho\eta) g$ evaluated at $\eta^{-1}{X}^\top{\rm P}_\mathcal{D}^\top\{{\rm P}_\mathcal{D} (\log y - {X}{\beta}^{(t-1)})  - {\theta}^{(t-1)} - \rho^{-1}{\Gamma}^{(t-1)} \} + {\beta}^{(t-1)}$.
One can check that using \eqref{eq:beta_update_approx},  $\mathcal{F}_\rho({\theta}^{(t-1)}, \beta^{(t)}, {\Gamma}^{(t-1)}) \leq \mathcal{F}_\rho({\theta}^{(t-1)}, \beta^{(t-1)}, {\Gamma}^{(t-1)})$ based on the majorize-minimize principle \citep{hunter2004tutorial}. This approximation was studied in Deng and Yin \cite{deng2016global}, who called this type of algorithm a ``prox-linear'' ADMM algorithm. From \eqref{eq:beta_update_approx}, one can see that our algorithm can be used for any estimator \eqref{eq:penalizedGehan} where the proximal operator of $g$ can be computed efficiently. Beyond the many penalties with closed form proximal operators, more sophisticated convex penalties (e.g., the overlapping group lasso \citep{yuan2011efficient} or fused lasso \citep{tibshirani2005sparsity}) have proximal operators which can be computed using efficient iterative algorithms.

Letting $\|A\|$ denote the spectral norm of a matrix $A$, we can summarize our proposed ADMM algorithm in Algorithm \ref{alg1}. 
This variation of the ADMM algorithm, which replaces the objective function in \eqref{eq:beta_update} with a quadratic approximation constructed at the previous iterate, is guaranteed to converge under reasonable conditions. 
\begin{prop}
Assume $g$ is convex and that a solution to \eqref{eq:penalizedGehan} exists.  If $\rho > 0$, $\lambda > 0$, $\eta  \geq \|X^\top{\rm P}_\mathcal{D}^\top{\rm P}_\mathcal{D}X\|$, and $\tau \in (0, (1 + \sqrt{5})/2)$, then as $t \to \infty$, the iterates $(\beta^{(t)}, {\theta}^{(t)}, {\Gamma}^{(t)})$ converge to $(\beta^{(\star)}, {\theta}^{(\star)}, {\Gamma}^{(\star)})$, where $(\beta^{(\star)}, {\theta}^{(\star)})$ are an optimal solution to \eqref{eq:Eff_ConstrainedProb} and  ${\Gamma}^{(\star)}$ is an optimal solution to the dual problem of \eqref{eq:Eff_ConstrainedProb}. 
\end{prop}
The proof of this result follows an identical argument as the proof Theorem 1 of Gu et al \cite{gu2018admm} and Theorem 2.2 of Deng and Yin \cite{deng2016global}.

\begin{algorithm}[t]\caption{Prox-linear ADMM algorithm for computing regularized Gehan estimator}\label{alg1}
   \setstretch{1.0}
    Initialize $(\beta^{(0)}, \theta^{(0)}, \Gamma^{(0)}) \in \mathbb{R}^p \times \mathbb{R}^{|\mathcal{D}|} \times \mathbb{R}^{|\mathcal{D}|}$, $\rho > 0$, $\eta \geq \|X^\top{\rm P}_\mathcal{D}^\top{\rm P}_\mathcal{D}X\|$, $\tau \in (0, (1 + \sqrt{5})/2)$, ${\Omega}^{(0)} = {\rm P}_\mathcal{D}(\log {y} - {X}{\beta}^{(0)})$ and set $t = 1$.
    %\vspace{-5pt}
\begin{itemize}
    \item[]\textit{1.} Compute ${\beta}^{(t)} = {\rm Prox}_{(\lambda/\rho\eta) g}\left\{ \eta^{-1}{X}^\top{\rm P}_\mathcal{D}^\top\left({\Omega}^{(t-1)} - 
    \rho^{-1}{\Gamma}^{(t-1)} - {\theta}^{(t-1)}\right) + {\beta}^{(t-1)}\right\}$ 
    \item[]\textit{2.} Compute ${\Omega}^{(t)} = {\rm P}_\mathcal{D}(\log {y} - {X}{\beta}^{(t)})$
   \item[]\textit{3.} Compute ${\phi} = {\Omega}^{(t)} - 
    \rho^{-1}{\Gamma}^{(t-1)}$ 
    \item[] \textit{4.} For each $k \in [|\mathcal{D}|]$, compute
    \begin{itemize}
        \item[]  ${\theta}^{(t)}_k = \left\{ {\phi}_k + \left(\frac{\tilde{{\delta}}_{k,1}}{{\rho n^2}}\right) \right\}\mathbf{1}\left({\phi}_k < -\frac{\tilde{{\delta}}_{k,1}}{{\rho n^2}}\right) + \left\{ {\phi}_k - \left(\frac{\tilde{{\delta}}_{k,2}}{{\rho n^2}}\right) \right\} \mathbf{1}\left({\phi}_k > \frac{\tilde{{\delta}}_{k,2}}{\rho n^2}\right)$
    \end{itemize}
    \item[]\textit{5.} Compute ${\Gamma}^{(t)} = {\Gamma}^{(t-1)} + \tau \rho 
    \left({\theta}^{(t)} - {\Omega}^{(t)} \right)$ 
    \item[]\textit{6.} If not converged, set $t = t+1$ and return to \textit{1}.
\end{itemize}
\end{algorithm}

\subsection{Implementation details}\label{sec:Implementation}
We implement Algorithm \ref{alg1}, along with a set of auxiliary functions, in the R package \texttt{penAFT} which can be downloaded from \url{https://github.com/ajmolstad/penAFT} or the Comprehensive R Archive Network. In this section, we provide some important details about our implementation. 

Following Boyd et al \cite{boyd2011distributed}, we monitor the progress of the algorithm based on the dual and primal residuals:
$s^{(t)} = \rho \|X^\top{\rm P}_\mathcal{D}^\top({\theta}^{(t)} - {\theta}^{(t-1)})\|_2$ and $r^{(t)} = \|{\theta}^{(t)} - {\rm P}_\mathcal{D}(\log {y} - {X}{\beta}^{(t)})\|_2,$ respectively.
We terminate the algorithm when $r^{(t)} < \epsilon_{\rm primal}^{(t)}$ and $s^{(t)} < \epsilon_{\rm dual}^{(t)}$ where, given the absolute and relative convergence tolerances $\epsilon_{\rm abs} > 0$ and $\epsilon_{\rm rel} > 0$, $\epsilon_{\rm primal}^{(t)} = \epsilon_{\rm abs}  \sqrt{|\mathcal{D}|} + \epsilon_{\rm rel} \max \{ \|{\rm P}_\mathcal{D}{X\beta}^{(t)}\|_2, \|{\theta}^{(t)}\|_2,  \|{\rm P}_\mathcal{D} \log {y}\|_2 \}$ and 
$\epsilon_{\rm dual}^{(t)} = \epsilon_{\rm abs}  \sqrt{p} + \epsilon_{\rm rel} \|X^\top{\rm P}_{\mathcal{D}}^\top{\Gamma}^{(t)}\|_2.$
In our package, we set $\epsilon_{\rm abs} = 10^{-8}$ and $\epsilon_{\rm rel} = 2.5 \cdot 10^{-4}$ as defaults, although a larger $\epsilon_{\rm rel}$ (e.g., $5 \cdot 10^{-4}$) is often sufficient when $p$ is large, 

The convergence of ADMM algorithms in practice is known to depend in part on the choice of step size parameter $\rho$. We intialize $\rho = 0.1$, which worked best amongst a number of values we tried. In Boyd et al \cite{boyd2011distributed}, an adaptive step size adjustment procedure is recommended at each iteration. However, we found this led to instability in certain instances. Instead, following Zhu \cite{zhu2017augmented}, we update the step size less frequently and incorporate the convergence tolerances. Step size updates occur at iterations $\{ \lfloor l_k \rfloor\}_{k=1}^\infty$ where, we first set $l_1 = 1$ and set $l_k = 1.1(l_{k-1}+1)$ for $k = 2, 3, 4, \dots$. Loosely, for iterations 1--14, step sizes are updated every other iteration; for iterations 15--26, step sizes are updated every third iteration, and so on. By the 250th iteration, the step size is updated approximately every thirty iterations. When updating the step size, we replace $\rho$ with  $2\rho$ if $r^{(t)}/\epsilon_{\rm primal}^{(t)} > 10 s^{(t)}/\epsilon_{\rm dual}^{(t)}$, replace $\rho$ with  $\rho/2$ if $s^{(t)}/\epsilon_{\rm dual}^{(t)} > 10 r^{(t)}/\epsilon_{\rm primal}^{(t)}$, and we leave $\rho$ unchanged otherwise. 
% $$ \rho = \left\{ \begin{array}{ll}
% 2\rho & :r^{(t+1)}/\epsilon_{\rm primal}^{(t+1)} > 10 s^{(t+1)}/\epsilon_{\rm dual}^{(t+1)} \\
% \rho/2 & :s^{(t+1)}/\epsilon_{\rm dual}^{(t+1)} > 10 r^{(t+1)}/\epsilon_{\rm primal}^{(t+1)}\\
% \rho & :\text{otherwise}
% \end{array}\right..$$
Like Zhu \cite{zhu2017augmented}, we found that incorporating the primal and dual convergence tolerances often led to faster convergence than the approach suggested in Boyd et al \cite{boyd2011distributed}.

% We found that for large $n$, it could occur that the iterates of $\boldsymbol{\theta}$ changed infrequently in early iterations, which could lead to slower convergence. To address this issue, we introduced an additional parameter, $\gamma \in [0,2]$, which we call the ``rescaling parameter". Specifically, our implementation is actually minimizing a scaled version of \eqref{eq:penalizedGehan}
% $$
% \argmin_{\boldsymbol{\beta} \in \mathbb{R}^p}  \left\{ \frac{1}{n^{2 - \gamma}} \sum_{i=1}^n \sum_{j=1}^n \delta_i \left\{ e_i(\boldsymbol{\beta}) - e_j(\boldsymbol{\beta}) \right\}^{-} + g_{\lambda n^{\gamma}}(\boldsymbol{\beta})\right\}
% .$$
% Evidently, this is the same objective function from \eqref{eq:penalizedGehan} multiplied by $n^{\gamma}$. By rescaling the objective function, the updating equation for $\boldsymbol{\theta}$ in Proposition 1 replaces all instances of $n^2$ with $n^{2 - \gamma}$ and the update for $\boldsymbol{\beta}$ requires solving the proximal operator of $g_{(\lambda n^\gamma)/(\rho\eta)}$. We set the balancing parameter equal to one in our implementation: we found that in general, this led to slightly faster convergence than did $\gamma = 0$.

In order to fit \eqref{eq:penalizedGehan} over a set of tuning parameters which yield relatively sparse models, our implementation determines a set of candidate tuning parameters for the user internally. Based on the Karush-Kuhn-Tucker (KKT) condition, ${\hat{\beta}_g}$ is an optimal solution to \eqref{eq:penalizedGehan} (with convex penalty $g$) if and only if
%\vspace{-10pt}
$$ 0 \in \frac{1}{n^2}\sum_{i=1}^n \sum_{j=1}^n \delta_i \left[ \partial \big\{e_i(\hat{\beta}_g) - e_j(\hat{\beta}_g)\big\}^{-} \right] + \lambda \partial g(\hat{\beta}_g) 
%\vspace{-10pt}
$$
where $\partial f(z)$ denotes the subdifferential of a function $f$ at $z$.  
% The subgradient of the Gehan loss function can be expressed 
% $$ \partial \{e_i(\beta) - e_j(\beta)\}^{-}  = (x_i - x_j) u_{i,j}(\beta), ~~~~~ u_{i,j}(\beta) = \left\{ 
% \begin{array}{ll}
% 1 & e_i(\beta) < e_j(\beta)\\
% 0 & e_i(\beta) > e_j(\beta)\\
% \left[0,1\right] & e_i(\beta) = e_j(\beta)
% \end{array}
% \right..
% $$
Letting $\partial f(\beta) = \frac{1}{n^2}\sum_{i,j} \delta_i [ \partial \{e_i(\beta) - e_j(\beta)\}^{-}] $ and $\mathcal{K} = \left\{(i,j): y_i = y_j, 1 \leq i \leq n, 1 \leq j \leq n \right\},$ one can verify that if $\lambda = \lambda_{\rm max}^{\rm EN}$ where
\begin{align*}
\lambda_{\rm max}^{\rm EN} & \geq \max_{1 \leq k \leq p} \left[ \frac{1}{n^2 \alpha w_k}\left\{ \left|\sum_{i=1}^n \sum_{j = 1}^n  \delta_i (x_{i,k} - x_{j,k})\mathbf{1}(y_i < y_j)\right| + \sum_{i = 1}^n \sum_{j=1}^n \mathbf{1}\{(i,j) \in \mathcal{K}\} \delta_i |x_{i,k} - x_{j,k}|\right\} \right],
\end{align*}
then ${\hat\beta} = 0$ under the elastic net penalty when $\alpha \neq 0$ and $w_k > 0$ for all $k \in [p]$. For the sparse group lasso, we can write the KKT condition as 
$$ 0 \in \partial f(\hat{\beta})+ \lambda \left(\alpha  \sum_{j=1}^p w_j \partial |\hat{\beta}_j| + (1 - \alpha) \sum_{l=1}^G v_l \partial\|\hat{\beta}_{\mathcal{G}_l}\|_2 \right),$$
so that letting ${\rm soft}(a, \tau) = \max(|a|-\tau, 0){\rm sign}(a)$,  $\hat{\beta} = 0$ is optimal (assuming all $w_k > 0$ and $v_l > 0$) if 
%\vspace{-10pt}
\begin{equation} \label{eq:KKT}
\|{\rm soft}(S_{\mathcal{G}_l}, \alpha \lambda w_{\mathcal{G}_l})\|_2 \leq v_l (1-\alpha) \lambda, \quad   S \in \partial f(0), \quad l \in [G].
%\vspace{-10pt}
\end{equation}
Hence, we attempt to find the minimum $\lambda$ such that the above holds. If $\mathcal{K} = \emptyset$, $\partial f(0)$ is a singleton, so that we can find $\lambda_{\rm max}^{\rm SG}$ using the fact that with $\alpha$ fixed, $\|{\rm soft}(S_{\mathcal{G}_l}, \alpha \lambda w_{\mathcal{G}_l})\|_2^2 - v_l^2 (1-\alpha)^2 \lambda^2$ is piecewise quadratic in $\lambda$ for each $l \in [G]$\citep{simon2013sparse}. If $\mathcal{K}$ is non-empty, we find a conservative $\lambda$ which guarantees a sparse solution, then compute the solution path until any coefficients become non-zero. Then, we set $\lambda_{\rm max}^{\rm SG}$ equal to the smallest tuning parameter value we considered which kept $\hat{\beta} = 0$. 

Once the $\lambda_{\rm max}$ has been computed, we construct the candidate tuning parameter set of length $M$, $\{\lambda_1, \dots, \lambda_M\}$ where $\lambda_i = 10^{\mu_i}$ for $i \in [M]$ where for some user-specified $\kappa \in (0,1)$, $\mu = \{\log_{10} \lambda_{\max}, \dots, \log_{10} (\kappa \lambda_{\max})\}$ consists of $M$ equally spaced points. In addition, to improve computational efficiency, we compute the entire solution path using ``warm-starting''. That is, we initialize the prox-linear ADMM algorithm for \eqref{eq:penalizedGehan} with $m$th largest tuning parameter $\lambda_{m}$ at the optimal values for \eqref{eq:penalizedGehan} with tuning parameter $\lambda_{m-1}$ for $m \in \{2, \dots, M\}.$ Since $\beta = 0$ is optimal for $\lambda_{1}$ by construction (when $\alpha \neq 0$ and the $w_k > 0$), we can use the KKT condition for \eqref{eq:Efficient_constrainedProb} to determine optimal initializing values for $\theta$ and $\Gamma$.

%\vspace{-10pt}
\subsection{Scalability}
Finally, we comment briefly on the computational complexity of our algorithm. When $n$ is large, a naive calculation involving quantities like ${\rm P}_\mathcal{D}X\beta$, may be problematic. To ensure efficiency, we first multiply and store ${X}{\beta}$, an $O(np)$ operation. Then, we multiply this $n$-dimensional vector by ${\rm P}_\mathcal{D} \in \mathbb{R}^{|\mathcal{D}| \times n}$. Considering that ${\rm P}_\mathcal{D}$ is extremely sparse (each of its $|\mathcal{D}|$ rows has only two non-zero entries) the multiplication of ${\rm P}_\mathcal{D}$ with ${X}{\beta}$ is $O(|\mathcal{D}|)$ when ${\rm P}_\mathcal{D}$ is stored as a sparse matrix. Of course, $|\mathcal{D}| \leq n(n-1)/2$ with equality only in the (worst) case where $\delta_i = 1$ for $i \in [n].$

%  In our implementation of Algorithm 1, which is in C++, we perform these types of matrix multiplications using for-loops: we found this to be faster than sparse matrix multiplications. 
%Moreover, multiplications like ${\rm P}_\mathcal{D}'{A}$ where ${A} \in \mathbb{R}^{|\mathcal{D}|}$ have complexity $O(n \sum_{i=1}^n \delta_i)$ as each row of ${\rm P}_\mathcal{D}'$ has at most $\sum_{i=1}^n \delta_i$ nonzero entries.  
%\vspace{-10pt}
\section{A new tuning parameter selection criterion}\label{sec:TuningCriterion}
Using penalized estimators of the form \eqref{eq:penalizedGehan}
requires the selection of one or more tuning parameters. Tuning parameters are often chosen by cross-validation, which requires the choice of a performance metric. In this section, we propose a new performance metric inspired by that of Dai and Breheny \cite{dai2019cross}, who studied various approaches for tuning parameter selection when fitting Cox proportional hazards models. Let $\mathcal{V}_1, \dots, \mathcal{V}_K$ be a random $K$ element partition of $[n]$ (the subjects) with the cardinality of each $\mathcal{V}_k$ (the $k$th fold) approximately equal for each $k \in [K]$. 
Let ${\hat{\beta}}_{\lambda(-\mathcal{V}_k)}$ be the solution to \eqref{eq:penalizedGehan} with tuning parameter $\lambda$ using only data indexed by $[n] \setminus \{\mathcal{V}_k\}$ (i.e., all but the $k$th fold). Previous works\cite{cai2009regularized,johnson2009rank} selected the tuning parameter according to
%\vspace{-10pt}
\begin{equation}\label{eq:deficientCV}
 \argmin_{\lambda \in \mathsf{\Lambda}} \sum_{k=1}^K \left[\frac{1}{|\mathcal{V}_k|^2}  \sum_{i \in \mathcal{V}_k} \sum_{j \in \mathcal{V}_k} \delta_i \{e_i({\hat{\beta}}_{\lambda(-\mathcal{V}_k)}) - e_{j}({\hat{\beta}}_{\lambda(-\mathcal{V}_k)})\}^{-}\right],
 %\vspace{-10pt}
 \end{equation}
where $\mathsf{\Lambda}$ is a user specified  (discrete) set of candidate tuning parameters. We refer to the value of this criterion at $\lambda$ as the cross-validated Gehan loss at $\lambda$. This approach, however, does not allow for leave-one-out cross-validation (i.e., $K = n$) since the criterion necessarily requires comparing $e_i$ and $e_j$ for some particular ${\hat{\beta}}_{\lambda(-\mathcal{V}_k)}$. Moreover, if the censoring proportion is high and $K$ is large, some folds will contain few subjects with observed failure times, in which case we observed \eqref{eq:deficientCV} to perform poorly.

Instead, we propose to use a criterion wherein the Gehan loss is not evaluated on each fold separately. Specifically, letting
$
\tilde{e}_i({\hat{\beta}}_\lambda) =  \sum_{k = 1}^K (\log y_i - x_i^\top{\hat{\beta}}_{\lambda(-\mathcal{V}_k)}) \mathbf{1}(i \in \mathcal{V}_k)$ for $i \in [n]$,
we choose $\lambda$ according to
%\vspace{-10pt}
$$
\argmin_{\lambda \in \mathsf{\Lambda}} \left[\sum_{i = 1}^n \sum_{j = 1}^n \delta_i \{\tilde{e}_i({\hat{\beta}}_\lambda) - \tilde{e}_j({\hat{\beta}}_\lambda)\}^{-}\right].
$$
Of course, by construction $i \in \mathcal{V}_k$ for only a single $k \in [K]$, so $\tilde{e}_i(\hat\beta_\lambda) = e_i(\hat\beta_{\lambda(-\mathcal{V}_k)})$ for all $i \in \mathcal{V}_k$.
This criterion, in contrast to \eqref{eq:deficientCV}, can be used for leave-one-out cross-validation, and performs well even if any $\mathcal{V}_{k}$ contains few subjects with observed failure times. This approach has been used for other models where model performance criterion cannot be evaluated on subsets of the data\cite{simon2011using}. Adopting the terminology from Dai and Breheny\cite{dai2019cross}, we refer to the values of this criterion as the cross-validated linear predictor score at $\lambda$. 
%\vspace{-20pt}

\section{Computing time experiments}
%\vspace{-10pt}
\subsection{Overview}\label{sec:CompTimeOverview}
In this section, we first present solution path computing times under both elastic net and sparse group lasso penalties. Then, we compare the solution path computing time of our method to that of the algorithms from Cai et al \cite{cai2009regularized} and Johnson \cite{johnson2009rank} under $L_1$-norm penalization.  Throughout, we use variations of the following data generating model. We assume that for a given $(n,p)$, each $x_i$ is a realization of ${\rm N}_p(0, \Sigma)$ where  $\Sigma_{j,k} = 0.5^{|j-k|}$ for $(j,k) \in [p] \times [p].$ Given $X = (x_1, \dots, x_n)^\top \in \mathbb{R}^{n \times p}$ and $\beta_* \in \mathbb{R}^p$, whose particular structure will be described separately, we generate failure times from the model 
$ \log T = X \beta_* + \epsilon$
where $\epsilon = (\epsilon_1, \dots, \epsilon_n)^\top$ with each $\epsilon_i$ being independent and identically distributed random variable following the logistic distribution with location parameter equal to zero and scale parameter equal to two. That is, ${\rm E}(\epsilon_i) = 0$ and ${\rm Var}(\epsilon_i) = 4 \pi^2/3$. Given a realization of $T = (T_1, \dots, T_n)^\top \in \mathbb{R}^{n}$, censoring times are drawn from an exponential distribution whose mean is equal to the 60th percentile of the $T_i$. All unspecified quantities ($\beta_*, n, p$) will be discussed in the next section. 

% \begin{figure}[!h]
% \begin{center}
% \includegraphics[width=15cm]{Plots/Timing_1.pdf}\\
% (a) Elastic net penalty  \\
% \smallskip
% \includegraphics[width=15cm]{Plots/Timing_2.pdf}\\
% (b) Sparse group lasso penalty
% \end{center}
% \caption{(a) Average computing times (in seconds) for the entire solution path (100 $\lambda$ values with $\kappa = 0.25$) under various alignments of $(n,p)$ with $\alpha \in \{0.50, 0.70, 0.80, 0.90, 0.95, 1\}$ using the elastic net penalty. (b) Average computing times (in seconds) for the entire solution  path (100 $\lambda$ values with $\kappa = 0.25$) under various alignments of $(n,p)$ with $\alpha \in \{0, 0.25, 0.50, 0.70, 0.80, 0.90\}$ using the sparse group lasso penalty. } \label{eq:ComputingTimes_1}
% \end{figure}

\begin{figure}[!h]
\begin{center}
\includegraphics[width=15cm]{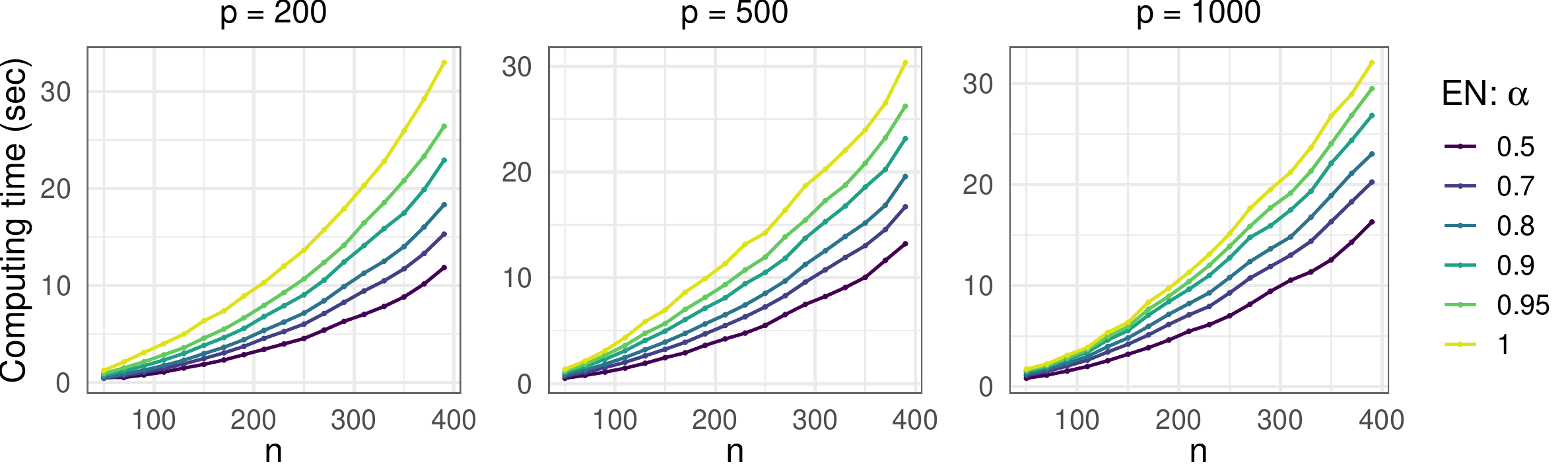}\\
(a) Elastic net penalty  \\
\smallskip
\includegraphics[width=15cm]{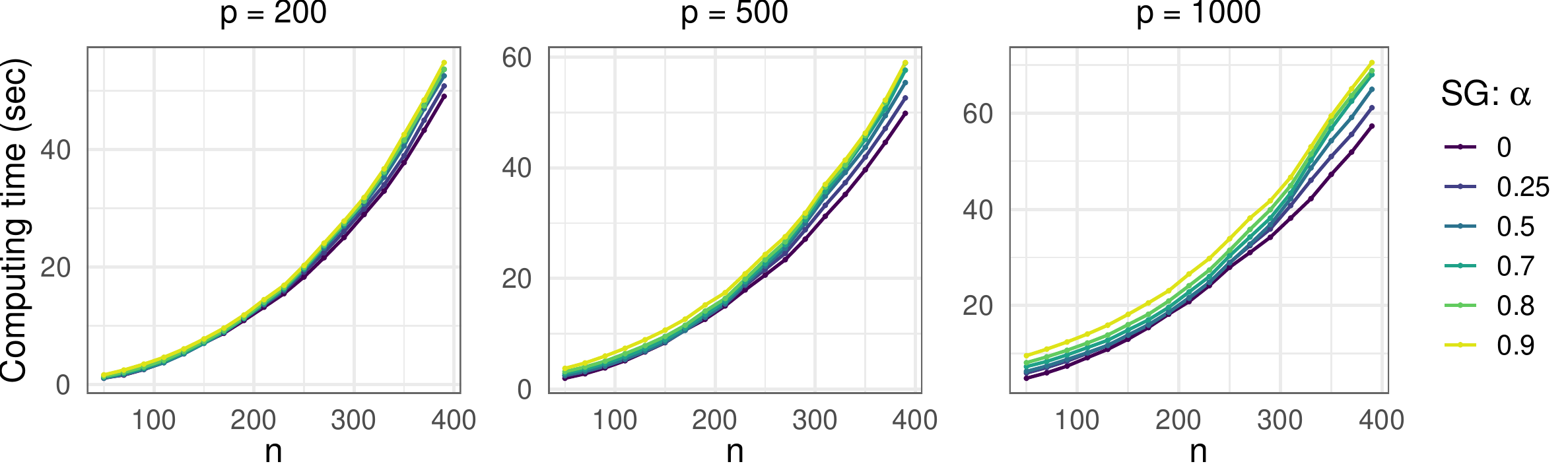}\\
(b) Sparse group lasso penalty
\end{center}
\caption{(a) Average computing times (in seconds) for the entire solution path (100 $\lambda$ values with $\kappa = 0.25$) under various alignments of $(n,p)$ with $\alpha \in \{0.50, 0.70, 0.80, 0.90, 0.95, 1\}$ using the elastic net penalty. (b) Average computing times (in seconds) for the entire solution  path (100 $\lambda$ values with $\kappa = 0.25$) under various alignments of $(n,p)$ with $\alpha \in \{0, 0.25, 0.50, 0.70, 0.80, 0.90\}$ using the sparse group lasso penalty. } \label{eq:ComputingTimes_1}
\end{figure}

\subsection{Solution path computing times}
We first assessed the time needed to compute the entire solution path on a single CPU for the elastic net penalized version of \eqref{eq:penalizedGehan} using our software. In each considered setting, we set $\beta_*$ to have ten randomly chosen entries equal to one and all others equal to zero. 
We considered $(n,p) \in \{50, 70, 90, \dots, 370, 390\} \times \{200, 500, 1000\}$. For 500 independent replications, we recorded the times needed to compute both the set of candidate tuning parameters with $\kappa = 0.25$ (see Section \ref{sec:Implementation}) and the entire solution path for the 100 candidate tuning parameter values. Convergence criteria were set at their default levels. We display results in the Figure \ref{eq:ComputingTimes_1}(a). As one may expect, as $\alpha$ approaches one, longer computing times were needed. This is because $\alpha < 1$ makes the objective function strongly convex: a smaller $\alpha$ means a larger strong convexity constant.

Next, we assessed the computing times for the sparse group penalized version of \eqref{eq:penalizedGehan}. Under the same settings as in Figure \ref{eq:ComputingTimes_1}(a), we divided the regression coefficients $\beta_*$ into $p/10$ groups: the first ten coefficients are one group, the second ten another group, and so on. We set the second group of ten coefficients all equal to $0.5$ and all others entirely equal to zero. 
Again considering $n \in \{50, 70, 90, \dots, 370, 390\}$ and $p\in \{200, 500, 1000\}$, for 500 independent replications, we recorded the times needed to compute both the set of candidate tuning parameters with $\kappa = 0.25$ and the entire solution path for 100 candidate tuning parameter values. We considered $\alpha \in \{0, 0.25, 0.50, 0.70, 0.80, 0.90\}$. Results are displayed in Figure \ref{eq:ComputingTimes_1}(b). Again, we see that as $n$ increases, computing times increased quadratically. Here, $\alpha$ does not control strong convexity, so it has a lesser effect.

\subsection{Computing time comparison to existing software}\label{subsec:compare_to_existing}
Next, we compared three different sets of software for fitting the $L_1$-penalized version of \eqref{eq:penalizedGehan} (i.e., elastic net with $\alpha = 1$). In addition to our own algorithm and software, we also used the algorithm based on the reformulation $h_M$, and the path-following algorithm of Cai et al \cite{cai2009regularized}. To implement the interior-point based approach of Johnson \cite{johnson2009rank}, we used the software downloaded from the author's webpage. For the path-following method of Cai et al \cite{cai2009regularized}, we used software provided by the authors through personal communication. For 500 independent replications under each considered setting, we first computed $\lambda_M = 0.5 \lambda_{\rm max}$. We then fit the solution path (terminating when $\lambda < \lambda_M$) using the method of Cai et al \cite{cai2009regularized}. Then, taking all tuning parameter values at which the path was evaluated by the method of Cai et al \cite{cai2009regularized}, we fit the path using both our software and the software of Johnson \cite{johnson2009rank}. 

\begin{figure}[!t]
\centering
\includegraphics[width=\textwidth]{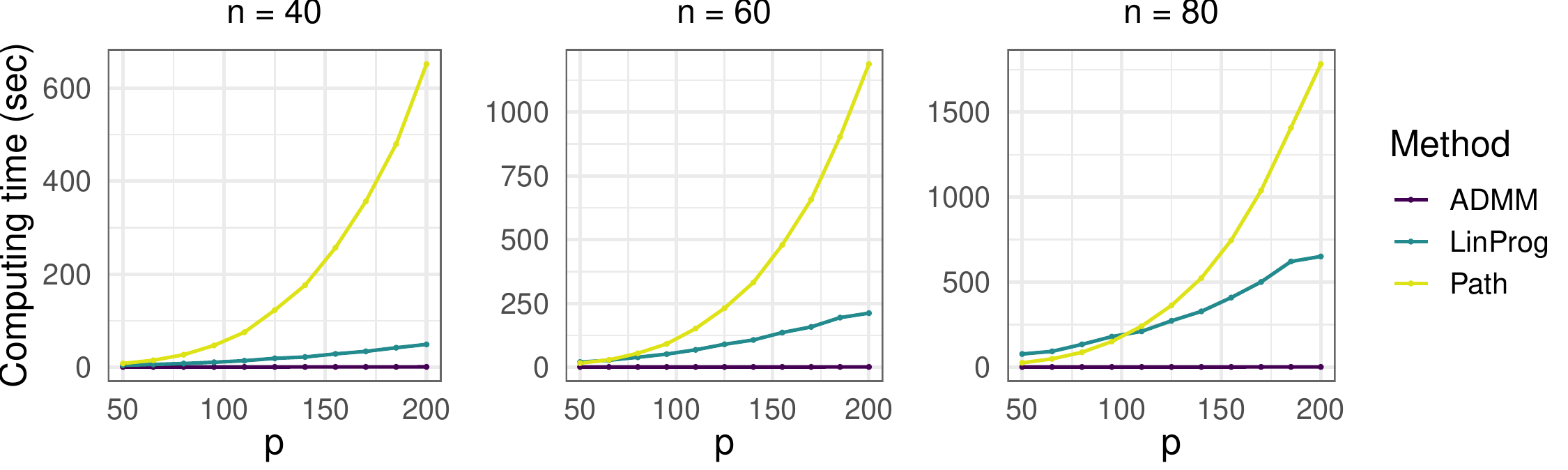}
\caption{Average computing times (in seconds) for the entire solution path using the path-based algorithm of Cai et al \cite{cai2009regularized} (\texttt{Path}), the interior point based approach of Johnson \cite{johnson2008estimation} (\texttt{LinProg}), and our proposed prox-linear ADMM algorithm and software (\texttt{ADMM}).  } \label{eq:ComputingTime_Comparisons}
\end{figure}

We see in Figure \ref{eq:ComputingTime_Comparisons} that the path-following method of Cai et al \cite{cai2009regularized} was slowest under almost every considered setting. The interior-point based method of Johnson \cite{johnson2009rank} was only slightly slower than our method when $n = 40$, but when $n = 80$ and $p = 200$, the difference was substantial. Accuracies for the three methods did not differ substantially. For example, in one replication with $n = 80$ and $p = 140$, the maximum difference in objective function values (which ranged from 1.276 at $\lambda_{\max}$ to 1.428 at $\lambda_{M}$) between our method and both competitors was $1.66 \cdot 10^{-5}$.

% \begin{figure}[!t]
% \centering
% \includegraphics[width=\textwidth]{Plots/TimingCompare.pdf}
% \caption{Average computing times (in seconds) for the entire solution path using the path-based algorithm of Cai et al \cite{cai2009regularized} (\texttt{Path}), the interior point based approximation of \citet{johnson2008estimation} (\texttt{LinProg}), and our proposed prox-linear ADMM algorithm and software (\texttt{ADMM}).  } \label{eq:ComputingTime_Comparisons}
% \end{figure}

It is important to note that these results are meant to compare the computing time of existing ``off-the-shelf'' software for obtaining the solution path of \eqref{eq:penalizedGehan}. Differences can partly be attributed to some factors beyond the efficiency of the respective algorithms. For example, our software is largely written in C++, whereas the code for the approach of Cai et al\cite{cai2009regularized} is written entirely in R. Similarly, the code implementing the algorithm from Johnson \cite{johnson2009rank} does not use warm-starting, so this implementation is not as efficient as one which is designed to compute the entire solution path as efficiently as possible.

For small scale settings like those in Figure \ref{eq:ComputingTime_Comparisons}, one could instead use \texttt{hqreg} to compute $\argmin_{\beta \in \mathbb{R}^p} h_M(\beta) + \lambda g(\beta)$. However, we found in larger scale settings (e.g., like those in Figure \ref{eq:ComputingTimes_1}), \texttt{hqreg} required much longer computing times than did our algorithm. In Section 4 of the  Supplementary Material, we provide a comparison of our method to the \texttt{hqreg} approach for solving $\argmin_{\beta \in \mathbb{R}^p} h_M(\beta) + \lambda g(\beta)$.  
% In an attempt to alleviate this issue, we tried using the \texttt{hqreg} package in R to minimize the $L_1$-penalized version of $h_M$, but found that the solutions could be highly unstable (e.g., objective function values 2 or 3 orders of magnitude greater than our algorithm) or, when requiring a more strict convergence tolerance, prohibitively time consuming (relative the computing times of the software of Johnson \cite{johnson2009rank}). This issue may partly explain why the loss $h_M$ has not been adopted as the ``standard'' computational approach for \eqref{eq:penalizedGehan}.
%This is likely due to the dominating effect of the appending large constant $M$ to $\log Y$ and $\sum_{i=1}^n \sum_{j=1} \delta_i (\mbx_i - \mbx_j)$ to $\mbX$. {}

%\vspace{-15pt}
\section{Simulation studies}
%\subsection{Overview}
In this section, we compare the regularized Gehan estimator \eqref{eq:penalizedGehan} to alternative estimators under the accelerated failure time model. In particular, we compare to variations of the regularized weighted least squares estimator of Huang et al \cite{huang2006regularized}
\begin{equation}\label{eq:stuteWLS} 
\argmin_{\beta \in \mathbb{R}^p} \left[ \left\{ \frac{1}{2n}\sum_{i=1}^n \xi_{i}(\log y_{(i)} - \beta^\top x_{(i)})^2\right\} + \lambda g(\beta)\right]
\end{equation}
where the $\xi_{i}$ are the jumps in the Kaplan-Meier estimator, $y_{(1)}, \dots, y_{(n)}$ are the order statistics for the $y_i$ (with $y_{(k)} \leq y_{(k+1)}$ for each $k$), and $(x_{(i)},\delta_{(i)})$ is the predictor and indicator of censoring, respectively, corresponding to $y_{(i)}$. Then, following Huang et al \cite{huang2006regularized}, the $\xi_{i}$ from \eqref{eq:stuteWLS} are defined as
$$ \xi_{1} = \frac{\delta_{(1)}}{n}, ~~~~ \xi_{i} = \frac{\delta_{(i)}}{n - i + 1} \prod_{j=1}^{i-1}\left(\frac{n-j}{n-j+1}\right)^{\delta_{(j)}}, ~~ i \in \{2, \dots, n\}.$$
While it has been shown that the unpenalized version of \eqref{eq:stuteWLS} was consistent (with $p$ fixed)\cite{stute1993consistent,stute1996distributional}, in finite samples \eqref{eq:stuteWLS} can perform poorly, especially in the case of high degrees of censoring. However, to compute \eqref{eq:stuteWLS} is straightforward using existing software (e.g., \texttt{glmnet}), so this method has been used widely in the literature. 

We consider four data generating models: the combination of two distributions for the $\epsilon_i$ and two structures for the $\beta_*$. For each scenario, we first generate the $x_i$ as realizations from ${\rm N}_p(0, \Sigma)$ where $\Sigma_{j,k} = 0.5^{|j-k|}$ for $(j,k) \in [p] \times [p].$ Then, we generate failure times using
$\log T = X \beta_* + \epsilon$
where $ \epsilon = (\epsilon_1, \dots, \epsilon_n)^\top$ with each $\epsilon_i$ either having (i) a logistic distribution with location parameter equal to zero and scale parameter $\sigma$ or (ii) having a normal distribution with mean zero and standard deviation $\sigma$. Censoring times are generated in the same manner as in Section \ref{sec:CompTimeOverview}.   We generate $n$ training samples and their censoring times; $200$ validation samples and their censoring times; and 1000 testing samples which are uncensored. 

To measure performance, we used concordance \citep{harrell1996multivariable} on the uncensored testing set (i.e., the degree of agreement in the ordering for all pairs of true survival times and linear predictors) and model error, which is defined as $(\hat\beta - \beta_*)^\top\Sigma(\hat\beta - \beta_*) = \lim_{n\to\infty} n^{-1}\|X\hat\beta - X\beta_*\|_2^2$ for random $X \in \mathbb{R}^{n \times p}$ generated as in Section \ref{sec:CompTimeOverview}.  In our data generating models ${\rm E}(\epsilon_i)= 0$, so ${\rm E}(\log T) = X\beta_*$ and thus, prediction of $X\beta_*$ is a reasonable goal. %Of course, a smaller model error indicates better performance.  Conversely, a concordance closer to one is better: a concordance of 0.5 indicates a model is no better than randomly guessing the order of survival times.  Both $n, p$, and $\sigma$ will vary across settings. 

To assess the performance of \eqref{eq:penalizedGehan} relative to \eqref{eq:stuteWLS} and the usefulness of the tuning parameter selection criterion described in Section \ref{sec:TuningCriterion}, we considered versions of \eqref{eq:penalizedGehan} with the tuning parameter selected by ten-fold cross-validation (\texttt{Gehan-CV(LP)}) and selected using a validation set (\texttt{Gehan-Val}). Note that $\texttt{Gehan-CV(LP)}$ does not use the validation set in any way, so in general, \texttt{Gehan-Val} has an advantage. Similarly, we consider selecting tuning parameters in \eqref{eq:stuteWLS} using both a validation set (\texttt{WLS-Val}) and based on ``oracle'' tuning (\texttt{WLS-Or}). For validation-set based estimators, we use tuning parameter which yields the lowest value of the Gehan loss function on the validation set. For the ``oracle'' estimator, we use the tuning parameter which had the smallest Gehan loss function value on the testing set -- an approach which could be not be used in practice. 

\begin{figure}[!t]
\centering
\includegraphics[width=\textwidth]{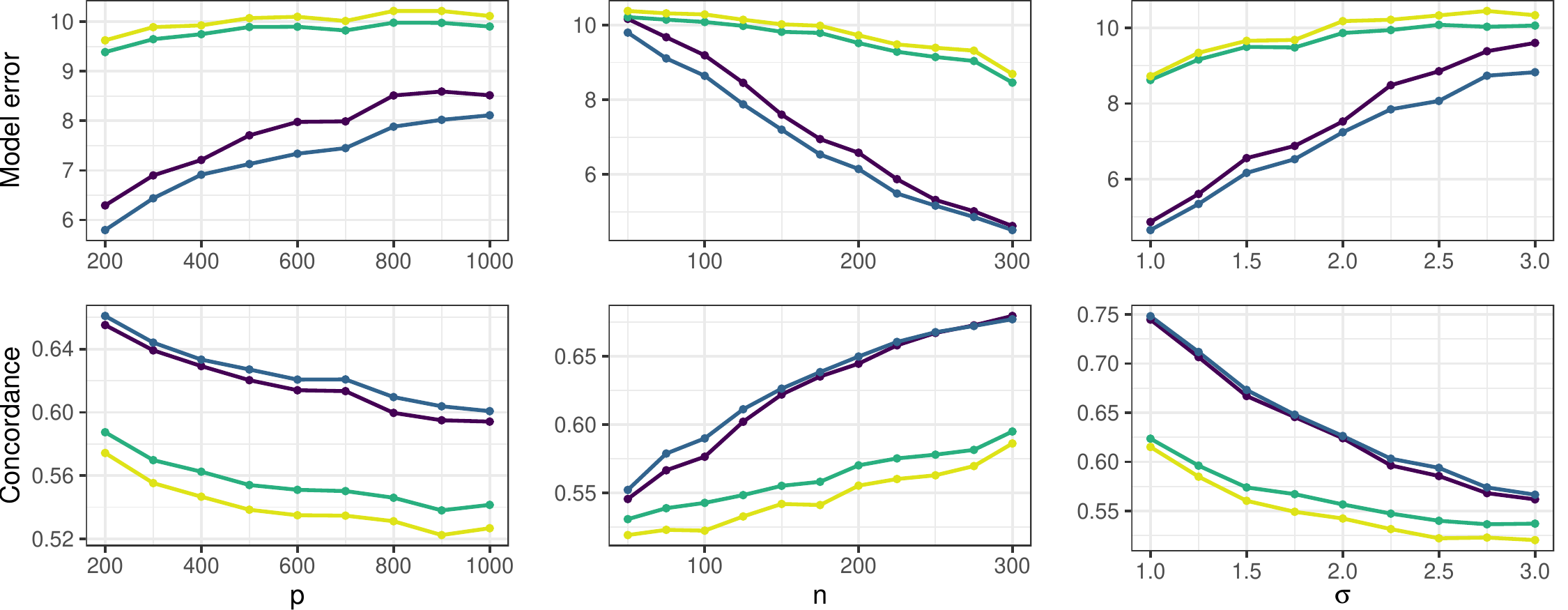}
\includegraphics[width=10cm]{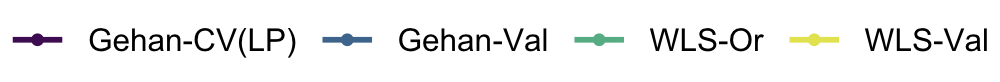}
\caption{Model error (top row) and concordance (bottom row) for the four considered methods averaged over 100 independent replications with logistic errors, $\beta_*$ having ten elements set equal to one, and $g$ being the elastic net penalty with $\alpha = 0.5.$ }\label{fig:EN_Sims_Logistic}
\end{figure}

\begin{figure}[!t]
\centering
\includegraphics[width=\textwidth]{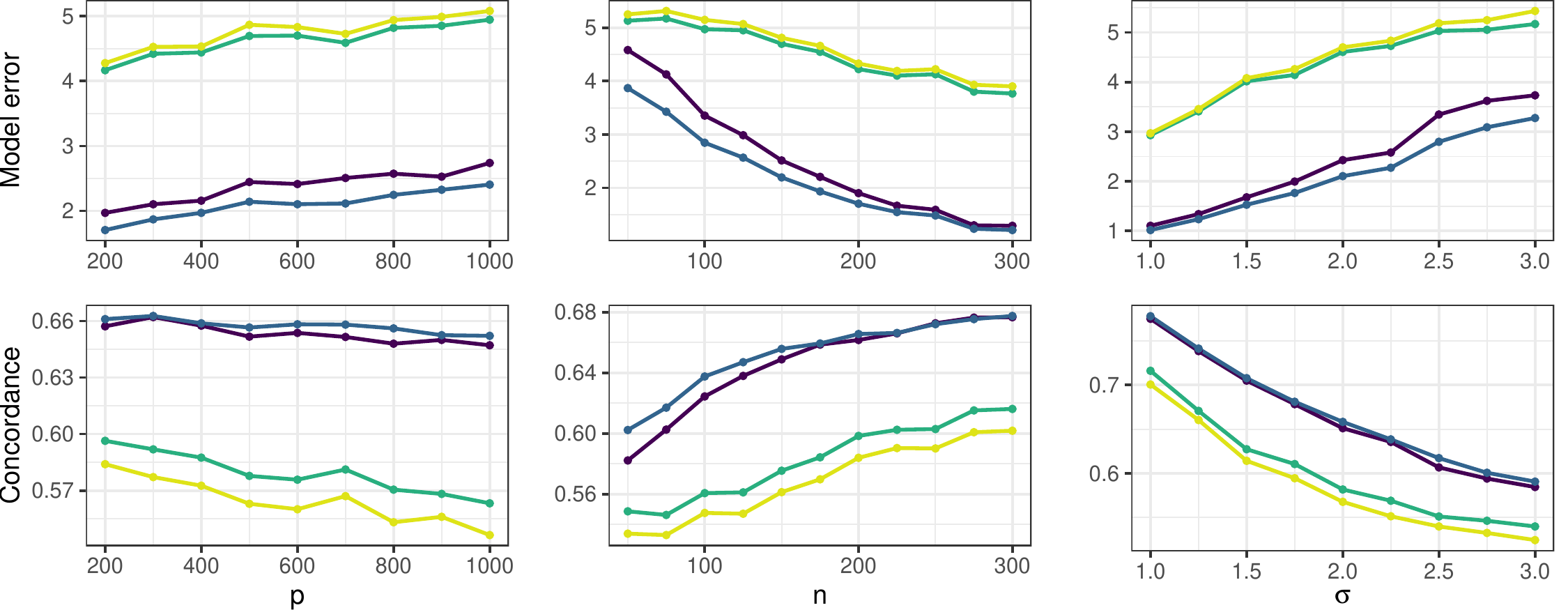}
\includegraphics[width=10cm]{Plots/Legend.png}
\caption{Model error (top row) and concordance (bottom row) for the four considered methods averaged over 100 independent replications with logistic errors, $\beta_*$ having ten elements set equal to 0.5 (five in two groups of size ten), and $g$ being the sparse group lasso penalty with $\alpha = 0.$ }\label{fig:Group_Sims_Logistic}
\end{figure}

%\subsection{Simulation studies}
We first compared \eqref{eq:penalizedGehan} with elastic net penalty to the elastic net penalized version of \eqref{eq:stuteWLS}. In these simulations, we constructed $\beta_*$ to have 10 randomly selected entries equal to one and all others equal to zero. For 100 independent replications, we considered $(n,p, \sigma) \in \{200\} \times \{100, 200, \dots, 800\} \times \{2\}$, $(n,p, \sigma) \in \{50, 100, 150, \dots, 450\} \times \{500\} \times \{2\}$, and $(n, p,\sigma) \in \{200\} \times \{500\} \times \{0.50, 0.75, \dots, 2.50\}$. 

Results with logistic and normal errors are displayed in Figure \ref{fig:EN_Sims_Logistic} and Figure 9 of the Supplementary Materials, respectively.  Across all the considered settings, both versions of \eqref{eq:penalizedGehan} outperformed the penalized weighted least squares estimator in both performance metrics.  Only when the sample size is very small (e.g., $n \approx 50$) or the noise level is very large (e.g., $\sigma \geq 3$) do we see the two sets of methods perform similarly in either metric. For logistic errors, this may be unsurprising given that least squares estimators are known to perform poorly with heavy-tailed errors. In the normal model, however, we still see that the rank-based estimators outperformed the weighted least squares estimator. We also performed these same simulations without censoring. Those results can be found in Figure 11 and 12 of the Supplementary Material. To summarize, in the case of normal errors, the penalized least squares approach (which is unweighted when there is no censoring) outperforms the regularized AFT model. Under logistic errors, the two methods perform nearly identically. %Together, these results suggest that when there is a reasonably high degree of censoring, rank-based estimators may be preferrable to the weighted least squares estimator in \eqref{eq:stuteWLS}.

In the next set of simulations, we compared performances of the two methods using the sparse group lasso penalty with $\alpha = 0$ (i.e., the group lasso penalty). In each replication, under the same model as before, we set $\beta_*$ to have $p/10$ groups of size ten. 
Each group has coefficients entirely equal to zero except the second (coefficients $11$ through $20$) and final group (coefficients $p-9$ through $p$), which have their first five coefficients equal to
0.5, and all others equal to zero. We use the same $(n,p, \sigma)$ configurations as in the elastic net setting. To compute regularized weighted least squares, we used the \texttt{gglasso} package in R. 

Averages over 100 independent replications are displayed in Figure \ref{fig:Group_Sims_Logistic} and Figure 10 of the Supplementary Materials. Just as in the elastic net simulations, \eqref{eq:penalizedGehan} outperformed the penalized weighted least squares estimator in nearly every considered setting. Notably -- and this applies to the elastic net case as well -- these are the best case versions of the weighted least squares estimator in the sense that one need not resort to cross-validation to select tuning parameters. Of course, this is not a feasible approach in practice.

Finally, we also measured variable selection performance using both true positive and true negative rates. Results under logistic errors are included in Figures 6--8 of the Supplementary Material. In brief, rank-based estimators tended to have much higher true positive rates and only slightly lower true negative rates. See Section 2 of the Supplementary Material for more details.

\section{TCGA data analyses}
%\vspace{-10pt}
\subsection{Comparison to weighted least squares and Cox model}
In our first real data application, we modeled survival as a function of gene expression (measured by RNAseq) in data collected from nine different cancer types by the Cancer Genome Atlas Project (Weinstein et al\cite{weinstein2013cancer}, https://portal.gdc.cancer.gov/). For each data type separately, we first performed screening by removing genes whose 75th percentile RNAseq count was less than 20. Then, we set the $i$th subject's $j$th gene expression equal to $\log_2\{(c_{i,j} + 1)/(q_{75,i})\}$ where $c_{i,j}$ is the sequencing count for the $i$th subject's $j$th gene and $q_{75,i}$ is the 75th percentile of counts for the $i$th subject across all genes. Finally, after these transformations, we kept only those genes with the $5000$ largest median absolute deviations across the entire dataset. We also included age as a predictor so that $p = 5001$. We did not impose any penalty on the coefficient corresponding to the patient's age.

In each dataset separately, for 100 independent replications, we randomly split the data into a training and testing set of sizes $n - \lfloor 0.2 n \rfloor$ and $\lfloor0.2 n \rfloor$, respectively. On the training data, we fit the $L_1$-penalized Cox model, the $L_1$-penalized version of \eqref{eq:penalizedGehan}, and the $L_1$-penalized version of the weighted least squares estimator of \eqref{eq:AFT_Model} \citep{huang2006regularized}. To measure model performance, we recorded both concordance (Harrell's C-index) and the integrated AUC measure proposed by Uno et al \cite{uno2007evaluating} (using the survAUC package in R) on the testing set. For a fair comparison between \eqref{eq:penalizedGehan} and the weighted least squares estimator of Huang et al \cite{huang2006regularized}, we used the same tuning parameter selection criterion -- that proposed in Section \ref{sec:TuningCriterion} based on 5-fold cross-validation -- for both methods. 

\begin{table}[t]\centering
\scalebox{.85}{
\fbox{
\begin{tabular}{c|cc|ccc|ccc|ccc}
  \multirow{2}{*}{Dataset} & \multirow{2}{*}{$n$} & \multirow{2}{*}{$\sum_{i=1}^n \delta_i$} & \multicolumn{3}{c|}{Concordance} &\multicolumn{3}{c|}{Integrated AUC} &\multicolumn{3}{c}{Computing time (secs)}  \\
&  &  & \texttt{penAFT} & \texttt{Cox} & \texttt{WLS} & \texttt{penAFT} & \texttt{Cox} & \texttt{WLS} & \texttt{penAFT} & \texttt{Cox} & \texttt{WLS} \\
  \hline
KIRC &530  & 174 & 0.713 & 0.718 & 0.559 & 0.747 & 0.751 & 0.564 & 209.6 & 17.3 & 24.7 \\ 
  LUAD & 480 & 173 & 0.612 & 0.578 & 0.562 & 0.599 & 0.562 & 0.555 & 230.8 & 20.2 & 24.2 \\ 
  LGG & 510 & 125 & 0.869 & 0.861 & 0.825 & 0.783 & 0.773 & 0.745 & 158.1 & 20.9 & 25.4 \\  
  LUSC & 489& 212& 0.534 & 0.532 & 0.524 & 0.512 & 0.516 & 0.506 & 235.3 & 22.4 & 24.4 \\ 
  BLCA & 404& 178& 0.653 & 0.638 & 0.601 & 0.657 & 0.640 & 0.593 & 164.1 & 16.7 & 24.1 \\ 
  KIRP & 283& 44 & 0.813 & 0.812 & 0.689 & 0.765 & 0.769 & 0.654 & 133.4 & 7.6 & 30.7 \\ 
  COAD & 277& 68 & 0.583 & 0.576 & 0.447 & 0.578 & 0.582 & 0.436 & 58.1 & 6.3 & 21.3 \\ 
  GBM & 151 &120 & 0.614 & 0.605 & 0.597 & 0.612 & 0.598 & 0.588 & 82.0 & 6.7 & 32.3 \\ 
  ACC & 79 & 28 & 0.834 & 0.842 & 0.710 & 0.826 & 0.836 & 0.723 & 17.2 & 1.5 & 18.4 \\ 
\end{tabular}
}
}
\caption{Average concordance, integrated AUC, and computing time for the three considered methods over the nine different cancer datasets from TCGA. Note that computing time includes the time taken for performing 5-fold cross-validtion and model fitting to the complete training dataset. For \texttt{penAFT}, we set $\epsilon_{\rm rel} = 5 \times 10^{-4}$ and left $\epsilon_{\rm abs}$ at its default value.}\label{tab:TCGA_Analysis}
\end{table}

Results are displayed in Table \ref{tab:TCGA_Analysis}. We see that the penalized Gehan estimator performed similarly to the method assuming proportional hazards (\texttt{Cox}) both in terms of concordance and integrated AUC. The weighted least squares approach of Huang et al \cite{huang2006regularized} performed comparatively worse. For example, in some datasets it had nearly 0.10 lower concordance than its competitors. Only in LUSC, where all methods predict only marginally better than random guessing, did we see the weighted least squares estimator perform similarly to \eqref{eq:penalizedGehan}. These results suggest that although the weighted least squares estimators are easy to implement, the ease of implementation comes with a potential sacrifice in predictive accuracy. The regularized Gehan estimator, on the other hand, may require slightly longer computing times (especially when $n$ is large), but yields fitted models which are competitive with estimators  assuming proportional hazards.

\subsection{Pathway-based analysis of KIRC data}
In this section, we return to the motivating pathway-based survival analysis described in Section \ref{sec:pathway_survival}. The goal was to fit the semiparametric accelerated failure time model treating genes belonging to particular pathways as a group. Specifically, we consider the six gene pathways used in Molstad et al \cite{Molstad2019Gaussian}. These are the (i) PI3K/AKT/mTOR pathway; four pathways associated with metabolic function: the (ii) glycolysis and gluconeogenesis, (iii) metabolism of fatty acids, (iv) pentose phosphate, and (v) citrate cycle pathways; and finally, (vi) the set of genes which were used in the CIBERSORT software (which we treat as a gene set). For a discussion of why these pathways are relevant to KIRC, see Section 5.2 of Molstad et al\cite{Molstad2019Gaussian} and references therein. As in the previous section, we also included age as a predictor: this is treated as its own group and was not penalized.

\begin{figure}[t]
\centering
\includegraphics[width=1\textwidth]{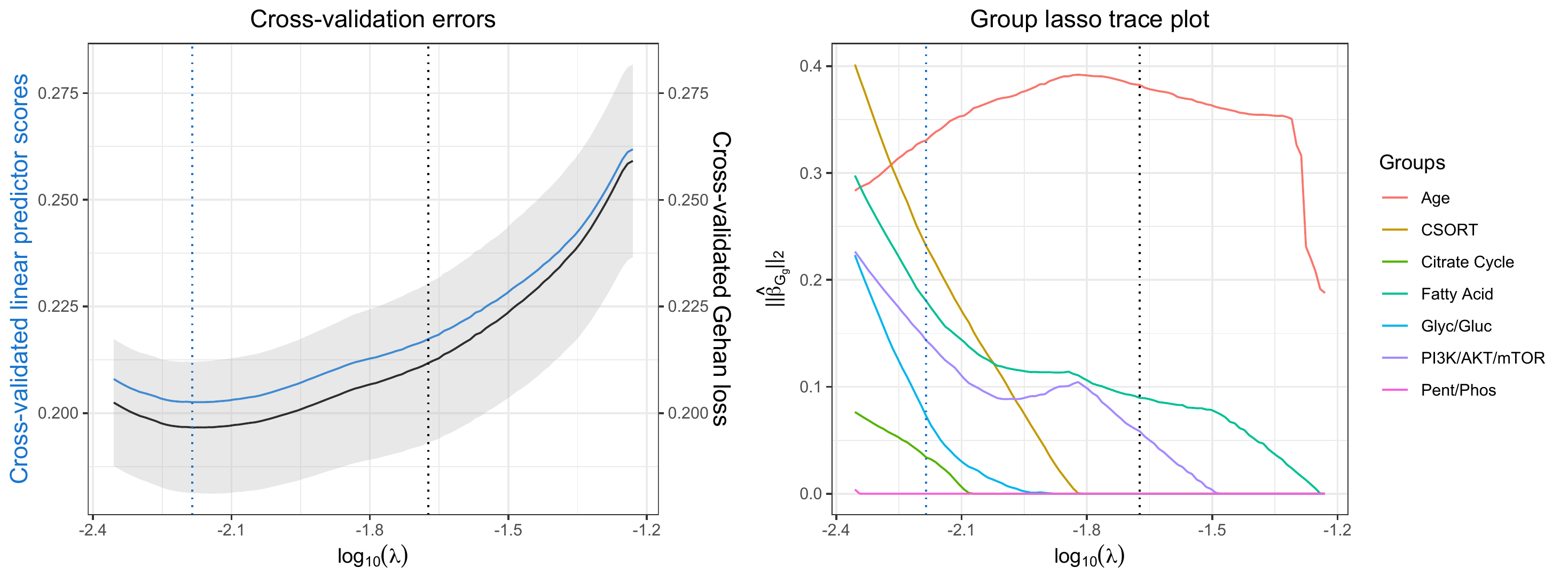}
\caption{(Left) Ten-fold cross-validation errors for the sparse group lasso pathway-based analysis of the KIRC dataset with $\alpha = 0$. The curve in blue is the cross-validated linear predictor scores (see Section \ref{sec:TuningCriterion}) and the black curve and standard error bands are the cross-validated Gehan loss \ref{eq:deficientCV}. Dotted vertical lines (blue) denote the tuning parameter with minimum cross-validated linear predictor score and (black) minimum cross-validated Gehan loss according to the one standard error rule. (Right) Trace plot for the sparse group lasso fit to the KIRC dataset with $\alpha = 0$.  }\label{eq:CV_curve}
\end{figure}

To perform this analysis, we fit \eqref{eq:sparseGroupLasso} to the full dataset. We included only those genes belonging to one of the six gene-sets, which leaves 581 genes for our analysis (so that $p = 582$). We set weights $v_g = \sqrt{p_g}$ where $p_g$ is the number of genes belonging to the $g$th group for $g \in [6]$. As before, the coefficient for age was not penalized. Because we assume that within certain gene-sets only a subset of genes may be needed, we considered $\alpha \in \{0, 0.05, 0.1, 0.25, 0.5\}$. 

First, we performed 10-fold cross-validation to select both $\lambda$ and $\alpha$. The minimum cross-validated linear predictor score was obtained with $\alpha = 0$, i.e., the group lasso without the $L_1$-penalty. The resulting cross-validation error curve is displayed in the left panel of Figure \ref{eq:CV_curve}. We saw that although the minimum cross-validated linear predictor score corresponds to a relatively large model, using the one standard error rule we would select a much smaller model. Looking at the corresponding trace plot displayed in the right panel of Figure \ref{eq:CV_curve}, we see that the model selected by the one standard error rule includes age, and all genes from both the fatty acid metabolism and PI3K/AKT/mTOR pathways. The model selected by the tuning parameter minimizing the cross-validated linear predictor score includes genes from all but the pentose phosphate pathway. It is not surprising that larger models may be preferable for this particular dataset: Molstad et al\cite{Molstad2019Gaussian} also found that larger models tended to outperform truly sparse models in another version of this dataset. 

\section{Discussion}
In this article, we have proposed a new algorithm for fitting semiparametric AFT models. We focused our attention on the elastic net and sparse group lasso penalties, but of course, the generality of our computational approach allows for a much wider range of penalties to be considered. Thus, our work affords practitioners a broad new class of estimators which were previously considered computationally infeasible. 

There are a number of interesting directions for future research. Specifically, the complexity of our algorithm scales quadratically in the number of subjects $n$. To allow for applications to data at the scale of the UK Biobank (https://www.ukbiobank.ac.uk/), which consists of roughly half a million subjects, new approaches need to be developed. One approach is to partition the subjects in the study into separate groups and apply \eqref{eq:penalizedGehan} on each group separately in a distributed fashion. However, this would require then combining the estimates in a theoretically justifiable way (e.g., as in Lee et al \cite{lee2017communication}), which is challenging since there is little in the way of theoretical studies of \eqref{eq:penalizedGehan} in high-dimensional settings. Alternatively, it may be preferable to devise and implement a version of our algorithm (or another ADMM variant) which is designed for parallelized, GPU-based computation. 

\subsection*{Acknowledgments}
The authors thank the associate editor and three referees for their helpful comments. The authors also thank Dr.\ Lu Tian for providing the code to implement the path-based algorithm from Cai et al \cite{cai2009regularized}, and thank Dr.\ Ben Sherwood for a helpful conversation. P.M. Suder's contributions were partially supported by the CLAS Scholars scholarship from the  College of Liberal Arts and Sciences at the University of Florida. A. J. Molstad's contributions were supported in part by a grant from the National Science Foundation (DMS-2113589).

\subsection*{Supplementary Material}
In the online Supplementary Material, we provide a proof of Lemma 1, a derivation of \eqref{eq:beta_update_approx}, and all additional results discussed in Section 6. We also include the R package \texttt{penAFT} along with a user guide.

\subsection*{Data Availability Statement}
The data that support the findings of this study are openly available from The Cancer Genome Atlas project (TCGA) at \href{https://portal.gdc.cancer.gov/}{https://portal.gdc.cancer.gov/}, and were downloaded on January 26th, 2021. 

\bibliography{penalizedGehan}

\end{document}